\documentclass[10pt, prl, twocolumn, superscriptaddress, showpacs, aps]{revtex4-2}
\usepackage{amsmath, amsthm, amssymb, cases} 

\usepackage{graphicx, float, overpic}

\usepackage[italicdiff]{physics}

\usepackage{textcomp, mathrsfs, dsfont, bm, amsbsy, bbold, xcolor}

\usepackage[OT2,T1]{fontenc}
    \DeclareSymbolFont{cyrletters}{OT2}{wncyr}{m}{n}
    \DeclareMathSymbol{\Sha}{\mathalpha}{cyrletters}{"58}

\usepackage{amsthm}
    \makeatletter
    \def\thmheadbrackets#1#2#3{%
      \thmname{#1}\thmnumber{\@ifnotempty{#1}{ }\@upn{#2}}%
      \thmnote{ {\the\thm@notefont[#3]}}}
    \makeatother
    
    \newtheoremstyle{definition}
        {5pt}
        {5pt}
        {\itshape}
        {}
        {\bfseries}
        {:\newline\indent}
        {.5em}
        {\thmheadbrackets{#1}{#2}{#3}}
    \theoremstyle{definition}
    
    \newtheoremstyle{approximation}
        {5pt}
        {5pt}
        {\itshape}
        {}
        {\bfseries}
        {:}
        {.5em}
        {\thmheadbrackets{#1}{#2}{#3}}
    \theoremstyle{approximation}

    \newtheoremstyle{scenario}
        {5pt}
        {5pt}
        {\rmfamily}
        {}
        {\bfseries}
        {:}
        {.5em}
        {\thmheadbrackets{#1}{#2}{#3}}
    \theoremstyle{scenario}

    \newtheoremstyle{remark}
        {5pt}
        {5pt}
        {\normalfont}
        {}
        {\itshape}
        {:}
        {.5em}
        {}
    \theoremstyle{remark}
    
\newcommand{\iu}{{i\mkern1mu}}

\newcommand{\integers}{\mathds{Z}}

\newcommand{\Dope}{\hat{D}}

\newcommand{\Hope}{\hat{H}}

\newcommand{\aope}{\hat{a}}

\newcommand{\zope}{\hat{z}}

\newcommand{\pope}{\hat{p}}

\newcommand{\sigmaope}{\hat{\sigma}}

\newcommand{\omegaa}{\omega_\mathrm{a}}

\newcommand{\omegaR}{\omega_\mathrm{R}}
\newcommand{\omegaS}{\omega_\mathrm{S}}

\newcommand{\omegarep}{\omega_\mathrm{rep}}

\newcommand{\phiR}{\phi_\mathrm{R}}

\begin{document}
\title{High-Fidelity Raman Spin-Dependent Kicks in the Presence of Micromotion}

\author{Haonan Liu}
\email{haonan.liu@ionq.co}
\affiliation{IonQ, Inc., College Park, MD, USA}
\author{Varun D. Vaidya}
\affiliation{IonQ, Inc., College Park, MD, USA}
\author{Monica Gutierrez Galan}
\affiliation{IonQ, Inc., College Park, MD, USA}
\author{Alexander K. Ratcliffe}
\affiliation{IonQ, Inc., College Park, MD, USA}%
\author{Amrit Poudel}
\affiliation{IonQ, Inc., College Park, MD, USA}
\author{C. Ricardo Viteri}
\affiliation{IonQ, Inc., College Park, MD, USA}

\date{\today}

\begin{abstract}
We propose high-fidelity single-qubit spin-dependent kicks (SDKs) for trapped ions using nanosecond Raman pulses via amplitude modulation of a continuous-wave laser with a tunable beat frequency. We develop a general method for maintaining SDK performance in the presence of micromotion by identifying optimal choices of the RF phase and frequency that suppress unwanted backward kicks. The proposed scheme enables SDK infidelities as low as $10^{-9}$ in the absence of micromotion, and below $10^{-5}$ with micromotion. This study lays the foundation for the realization of sub-trap-period and high-fidelity two-qubit gates based on SDKs.
\end{abstract}

\maketitle

Among the competing quantum computing hardware platforms, trapped ions stand out for their long coherence times, high-fidelity gates, and reconfigurable all-to-all connectivity~\cite{blatt_entangled_2008, bruzewicz_trapped-ion_2019}. Scaling up trapped-ion processors, however, remains a central challenge. The prevailing architecture for achieving this scalability is the quantum charge-coupled device~(QCCD) model~\cite{kielpinski_architecture_2002, pino_demonstration_2021, akhtar_high-fidelity_2023, paetznick_demonstration_2024, mordini_multizone_2025, jones_architecting_2025}, where ions are shuttled between zones to perform gates. While QCCD enables modular design and reconfigurable connectivity, the repeated shuttling and re-cooling of ions significantly limit the overall operation speed~\cite{moses_race-track_2023, pino_demonstration_2021}. Alternatively, architectures based on long ion chains avoid shuttling but introduce a dense and complex motional-mode spectrum that constrains gate speed and fidelity~\cite{ debnath_demonstration_2016, figgatt_parallel_2019, wright_benchmarking_2019, grzesiak_efficient_2020, hou_individually_2024, chen_benchmarking_2024, schwerdt_scalable_2024}. 

To overcome these constraints, non-adiabatic two-qubit ``fast gates'' driven by impulsive spin-dependent kicks~(SDKs) have been proposed~\cite{garcia-ripoll_speed_2003, duan_scaling_2004, campbell_ultrafast_2010, bentley_fast_2013, gale_optimized_2020, ratcliffe_micromotion-enhanced_2020, an_programmable_2025}. These gates couple to the collective spin-dependent motion of the ion crystal, thereby circumventing the need to resolve dense motional spectra. Recent theoretical studies reveal that high-fidelity two-qubit gates can be designed in this regime while maintaining the direct all-to-all connectivity advantage of long ion chains~\cite{mehdi_fast_2021, mehdi_fast_2025, savill-brown_error-resilient_2025, savill-brown_high-speed_2025}. The concept has also been theoretically extended to operations in interconnected trapping potentials, offering a path to modularity without shuttling~\cite{ratcliffe_scaling_2018, mehdi_scalable_2020}. However, demonstrating high-fidelity two-qubit gates has proven difficult. Experimental realizations showed a fidelity of 76\%~\cite{wong-campos_demonstration_2017}, where the main contributor to the error budget was the~0.99 fidelity of each SDK~\cite{johnson_sensing_2015, johnson_ultrafast_2017}. The limited fidelity of SDKs stems from two primary sources: parasitic four-photon~(or higher-order) processes inherent in prior pulsed-laser schemes, and unaddressed micromotion effects, which occur on timescales comparable to a single SDK~\cite{wong-campos_demonstration_2017, ratcliffe_micromotion-enhanced_2020}.

\begin{figure}
    \centering
    \begin{minipage}{0.48\linewidth}
        \centering
        \begin{overpic}[width=\linewidth]{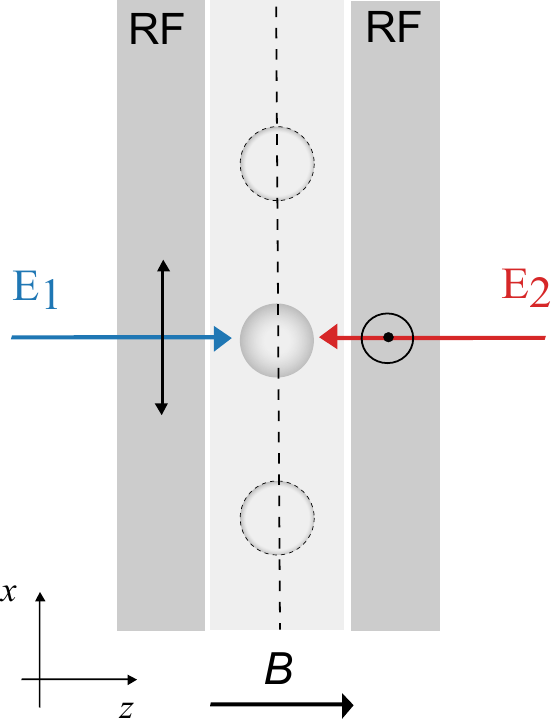}  
            \put(2,92){\textbf{(a)}}
        \end{overpic}
    \end{minipage}
    \begin{minipage}{0.41\linewidth}
        \centering
        \begin{overpic}[width=\linewidth]{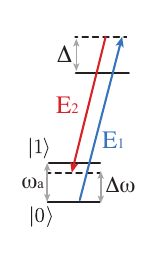}  
            \put(2,92){\textbf{(b)}}
        \end{overpic}
    \end{minipage}
    \caption{(a) Schematic of the Raman spin-dependent kick~(SDK). Two counterpropagating laser beams~($E_1$ and~$E_2$) with lin $\perp$ lin polarization interact with a single ion in a linear ion chain aligned along the $x$ axis (dashed line). The ions are confined in a linear Paul trap with RF (shaded) and DC (not shown) electrodes. The quantization axis is set by a static magnetic field $B$ along the $z$ axis. (b) A generic level diagram of the Raman transition. An ion initially in the spin ground state $\ket{0}$ experiences a forward kick when it absorbs a photon from $E_1$ and emits into $E_2$, gaining momentum in the~$+z$ direction. Here $\Delta$ denotes the detuning of the Raman beams from the excited states, $\omegaa$ is the qubit frequency splitting, and $\Delta\omega$ is the frequency difference between the two Raman beams.}
    \label{fig: 1}
\end{figure}

In this Letter, we resolve these challenges by developing a comprehensive SDK model in a continuous-wave~(CW) scheme that fully incorporates the effects of micromotion. Our model predicts infidelities as low as~$10^{-5}$ for nanosecond SDKs in experimentally accessible conditions. In contrast to previous pulsed schemes that implement a single nanosecond SDK using multiple picosecond pulses from mode-lock lasers~\cite{wong-campos_demonstration_2017, campbell_ultrafast_2010, mizrahi_ultrafast_2013_paper}, we introduce a CW scheme where each kick is realized by a single smooth nanosecond pulse that is shaped via modulators from a CW laser source. By shaping the pulse envelope and tuning the modulation parameters, we show that our SDKs can suppress multi-photon errors found in the previous pulsed scheme while requiring significantly lower peak optical power. Crucially, we incorporate the effects of intrinsic micromotion in the Paul trap, which can become non-negligible on nanosecond timescales. We analytically identify the conditions for micromotion phase matching, and verify them through full quantum simulations, revealing parameter regimes where micromotion-induced errors vanish. We further demonstrate that our optimized CW SDKs maintain high fidelity under realistic control errors and trap parameters, establishing a clear path toward robust, fast, and experimentally feasible SDK-based entangling gates. Our results provide physical intuition linking SDKs in the CW and pulsed schemes, showing that both can be understood within a common framework. This work thus advances the design of high-fidelity, fast entangling gates in trapped-ion quantum processors by bridging the gap between pulsed laser and continuous-wave based SDK implementations.

We consider a single ion with qubit states $\qty{\ket{0},\ket{1}}$ confined in a linear Paul trap and driven by two counter-propagating Raman beams~($E_1$ and~$E_2$) along the axis~$z$, which is defined by a static magnetic field~$B$~[Fig.~\ref{fig: 1}(a)]. Here, the Raman beams are arranged in a lin~$\perp$~lin polarization configuration~\footnote{For more details on this choice of configuration, see~\cite{campbell_ultrafast_2010}.} with wavevectors~$\pm \vb{k} = \qty(0, 0, \pm k)$, and the qubit states are encoded in the ion's hyperfine ground manifold. Throughout this work, we use the clock qubits in the~${6}^{2}\mathrm{S}_{1/2}$ manifold of~${}^{133}\mathrm{Ba}^{+}$ as an example, which provides a qubit splitting of~$\omegaa \approx 2\pi \times 10~\mathrm{GHz}$. However, the same scheme applies to other species with comparable level structures, such as~${}^{171}\mathrm{Yb}^{+}$. Figure~\ref{fig: 1}(b) shows a generic level diagram, where the two Raman beams, detuned from the excited state by~$\Delta$ and with an adjustable frequency difference~$\Delta\omega$, off-resonantly couple the qubit states separated by~$\omegaa$. The total time-dependent Hamiltonian of the system is then given by~\footnote{See Supplemental Material for details.}
\begin{align}
    \Hope(t) = &\frac{\pope_z^2}{2m} + 
           \frac{1}{8} m \omegaR^2 \zope^2 \qty[a_z + 2 q_z \cos\qty(\omegaR t + \phiR)] \notag\\
\label{eq: H_SP}
           & + \frac{\hbar\omegaa}{2} \sigmaope_z + 
           \hbar\Omega(t) \cos\qty(2 k \zope - \Delta\omega t) \sigmaope_x.
\end{align}
Here,~$\zope$ and~$\pope_z$ are the position and momentum operators of the ion along the~$z$ direction,~$a_z$ and~$q_z$ are the~$z$-direction Mathieu parameters,~$\omegaR$ and~$\phiR$ are the frequency and phase of the RF drive, and~$m$ is the ionic mass. The ion’s motion includes both secular motion and intrinsic micromotion components due to the RF trap drive. Operators~$\sigmaope_z$ and~$\sigmaope_x$ are the Pauli operators,~$\Omega(t)$ is the time-dependent two-photon Rabi frequency of the Raman transition, and~$\Delta\omega$ is the frequency difference, or the Raman beat frequency, of the two beams that is much smaller than the optical frequency. We also assume that the two beams have the same intensity profile and initial phase in our analysis. 

The physical picture of the spin-dependent force, which results in the SDKs, is manifest if we rewrite the Hamiltonian in Eq.~\eqref{eq: H_SP} in the interaction picture defined by~$\Hope_0 = \hbar\omegaS \aope^\dagger \aope + \hbar\omegaa \sigmaope_z / 2$, where $\omegaS = \omegaR \sqrt{a_z + \frac{q_z^2}{2}}/2 $ is the secular frequency, and $\aope$ and $\aope^\dagger$ are the lowering and raising operators corresponding to the secular motion of the ion, yielding~\footnote{See Supplemental Material for details.}
\begin{align}
    \tilde{\Hope}(t) = &\underbrace{\frac{\hbar}{16} \frac{\omegaR^2}{\omegaS} \qty(e^{\iu \omegaS t} \aope^\dag + e^{-\iu \omegaS t} \aope)^2 \qty[2 q_z \cos\qty(\omegaR t + \phiR) - \frac{q_z^2}{2}] }_{\mathrm{micromotion}} \notag\\ 
      & + \frac{\hbar\Omega(t)}{2} \Bigg[\underbrace{\qty(\tilde{\Dope}_+ \sigmaope_+ e^{\iu \qty(\omegaa - \Delta\omega) t} + \tilde{\Dope}_- \sigmaope_- e^{- \iu\qty(\omegaa - \Delta\omega) t})}_{\mathrm{forward\,kick}} \notag\\
\label{eq: H_IP}
      &+ \underbrace{\qty(\tilde{\Dope}_+ \sigmaope_- e^{-\iu \qty(\omegaa + \Delta\omega) t} + \tilde{\Dope}_- \sigmaope_+ e^{\iu\qty(\omegaa + \Delta\omega) t})}_{\mathrm{backward\,kick}}\Bigg],
\end{align}
where $\tilde{\Dope}_\pm(t) = \Dope(\pm 2\iu \eta e^{\iu \omegaS t}) = e^{\pm 2\iu \eta \qty(e^{\iu \omegaS t}\aope^\dagger + e^{- \iu \omegaS t} \aope)}$ are the displacement operators in the interaction picture, and $\eta = k\sqrt{\hbar/\qty(2m\omegaS)}$ is the Lamb-Dicke parameter.

\begin{figure*}[t]
    \centering
    \begin{minipage}{0.288\linewidth}
        \centering
        \begin{overpic}[width=\linewidth]{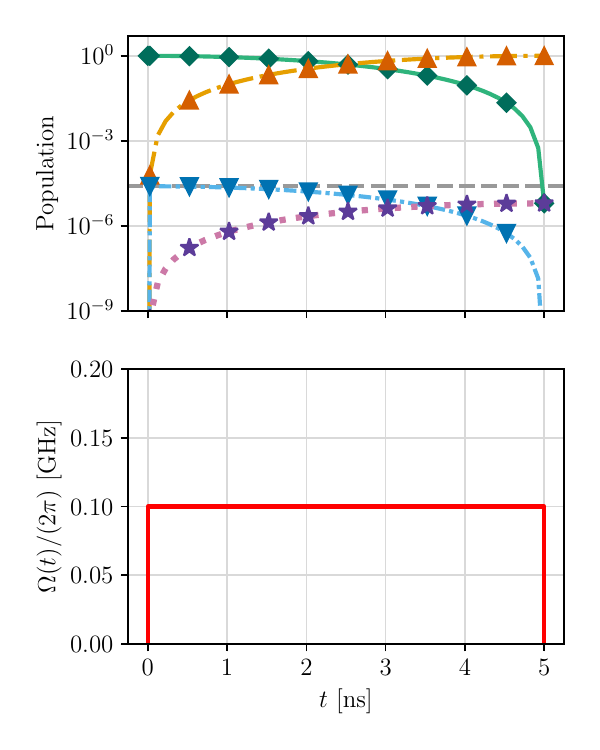}  
            \put(5,100){\textbf{(a)}}
        \end{overpic}
    \end{minipage}
    \begin{minipage}{0.288\linewidth}
        \centering
        \begin{overpic}[width=\linewidth]{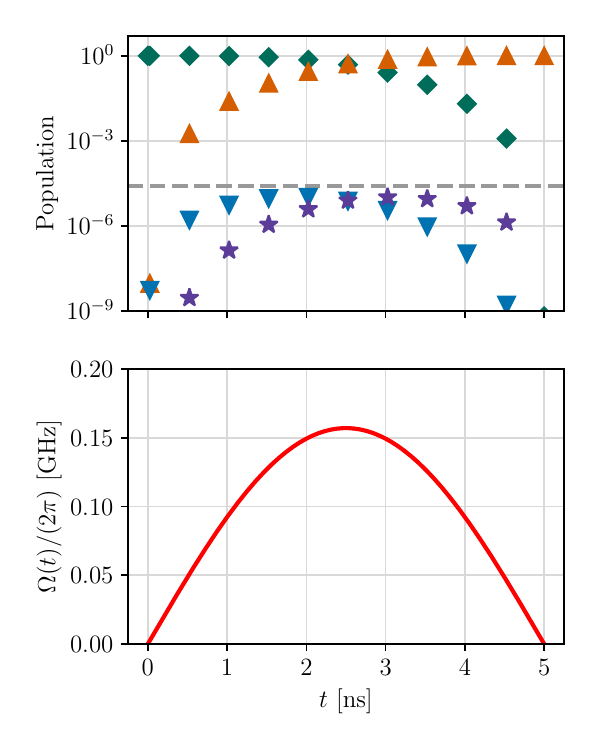}  
            \put(5,100){\textbf{(b)}}
        \end{overpic}
    \end{minipage}
    \begin{minipage}{0.288\linewidth}
        \centering
        \begin{overpic}[width=\linewidth]{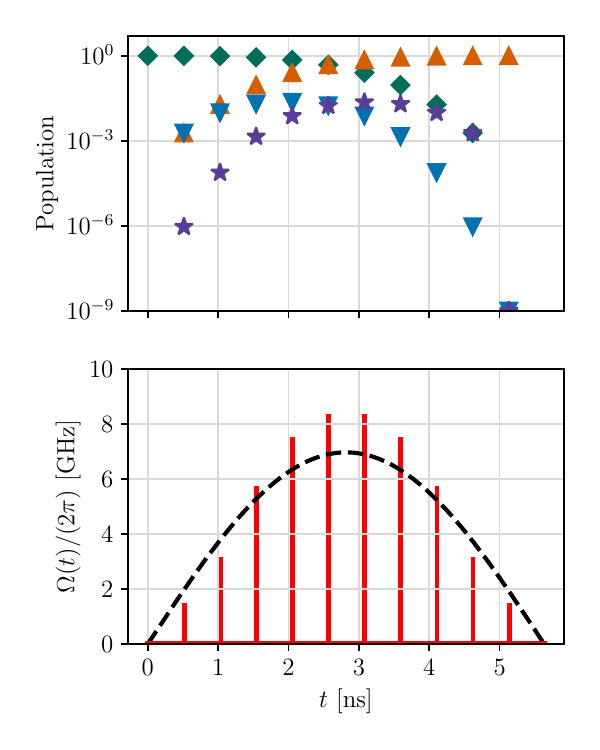}  
            \put(5,100){\textbf{(c)}}
        \end{overpic}
    \end{minipage}
    \begin{minipage}{0.12\linewidth}
        \centering
        \includegraphics[width=\linewidth]{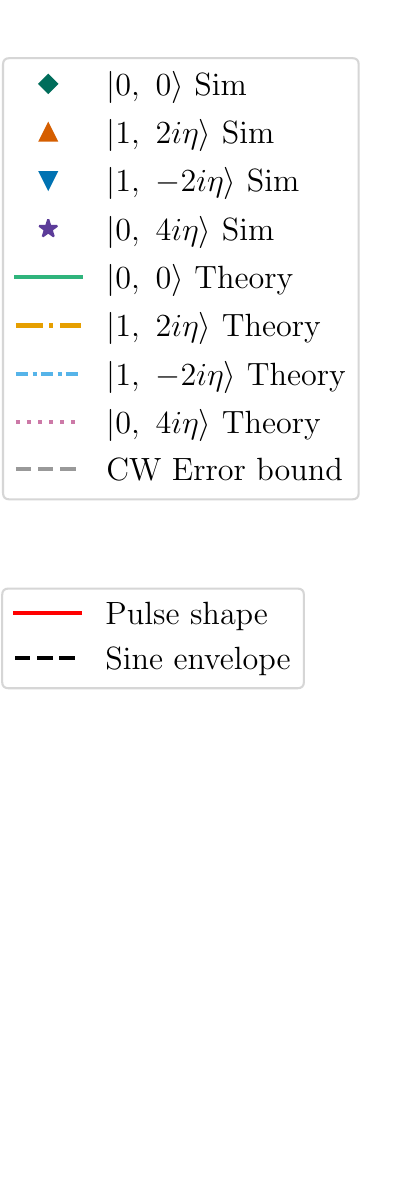}  
    \end{minipage}
    \caption{Population dynamics and pulse envelopes of continuous-wave (CW) and pulsed SDKs as a function of time without including micromotion. (a) For a~$5~\mathrm{ns}$ constant-amplitude pulse (bottom), both analytical and simulation calculations (top) yield an infidelity of~$1 - \mathcal{F} = 1.9 \times 10^{-5}$ at the exact resonant condition~$\Delta\omega = \omegaa$. (b) For a~$5~\mathrm{ns}$ sine-shaped pulse (bottom), the infidelity is reduced to~$1.4 \times 10^{-9}$ when the Raman beat frequency is optimized to~$\Delta\omega = \qty(1 + 5 \times 10^{-5}) \omegaa$. Both CW SDKs in (a) and (b) employ a total pulse area of~$\theta = \pi$ and a peak Rabi frequency~$\Omega_\mathrm{max} \gtrsim 2\pi \times 100~\mathrm{MHz}$. (c) Pulsed SDKs can achieve comparable~$10^{-9}$ infidelity by approximately sampling the sine envelope and optimizing the repetition rate~$\omegarep$ and Raman beat~$\Delta\omega$. For a sequence of ten $10~\mathrm{ps}$ pulses of optimized amplitudes within~$5.6~\mathrm{ns}$ (bottom), the infidelity of a single pulsed SDK is~$3.0 \times 10^{-9}$, with~$\Delta\omega = 0.027 \omegaa$ and~$\omegarep = 2\pi \times 1.9~\mathrm{GHz}$. The corresponding peak Rabi frequency exceeds~$2\pi \times 8~\mathrm{GHz}$.}
    \label{fig: 2}
\end{figure*}

The Hamiltonian in Eq.~\eqref{eq: H_IP} can be separated into three components that capture the essential physics of the SDK in the presence of micromotion. The first is the micromotion term, which arises from the RF-driven motion of the ion and is independent of the spin degree of freedom. This term introduces a time-dependent phase at the RF frequency~$\omegaR$ in the evolution, and its strength can be tuned through the trap’s Mathieu parameters. In conventional MS gates, the micromotion term is typically averaged out because its timescale is much faster than the gate duration, yielding the usual harmonic trap approximation~\cite{cook_quantum_1985, nguyen_micromotion_2012}. In our regime, however, the micromotion timescale is comparable to that of the SDK, and its effects must therefore be explicitly included for realistic high-fidelity design. The second component corresponds to the \textit{forward kick}, representing the resonant interaction in which an ion initially in the ground state~$\ket{0}$ absorbs a photon from $E_1$, emits a photon into $E_2$, and transitions to the excited state~$\ket{1}$ while acquiring a net momentum kick of $2\hbar k$ in the $+z$ direction (and vice versa if the ion starts in the excited state). This process drives the desired spin–motion entanglement, thus producing a spin-dependent force. The third component is the \textit{backward kick}, a counter-rotating term that drives the opposite spin-dependent transition. In the nanosecond-scale regime of the SDKs, all three components contribute to the overall dynamics. Our objective in this work is to suppress the unwanted backward kick and adverse micromotion effects by optimizing system parameters, including the RF drive~$\omegaR$, RF phase~$\phiR$, Rabi frequency~$\Omega(t)$, and Raman beat frequency~$\Delta\omega$~\footnote{Technically, the distinction between the ``forward'' and ``backward'' components is arbitrary; the essential requirement is to maximize one while suppressing the other.}.

Our approach proceeds in three stages to systematically isolate and understand the key contributions to the SDK dynamics. We first neglect the micromotion term and analyze CW SDKs driven by nanosecond-scale pulses. In this limit, the pulse envelope can be treated as effectively continuous, allowing us to suppress the backward (counter-rotating) kicks and achieve near-resonant spin–motion coupling. We then turn to the micromotion contribution alone and derive analytical conditions under which its effects vanish in the fast SDK regime, where the SDK duration $\tau$ is much less than the secular motion timescale, i.e.,
\begin{align}
\label{eq: fast_SDK_limit}
    \omegaS \tau \ll 1.
\end{align}
Finally, we combine both effects to identify the experimentally relevant parameter regime that simultaneously minimizes micromotion-induced errors and backward-kick contributions, enabling realistic, high-fidelity implementation of fast SDKs.

We begin by analyzing SDKs in the absence of micromotion. In this simplified regime, we set the Raman beat frequency to be exactly the qubit splitting, i.e.,~$\Delta\omega = \omegaa$. The SDK is implemented using a constant-amplitude pulse of area~$\theta = \pi$ and duration~$\tau = 5~\mathrm{ns}$, which can be regarded as a CW envelope relative to the Raman beat frequency. Then by Eq.~\eqref{eq: H_IP}, the backward kick term can be ignored in the rotating wave approximation with an analytical error bound~\footnote{See Supplemental Material for details.} of~$\qty[\theta/(2\omegaa \tau)]^2$. Fig.~\ref{fig: 2}(a) shows this error bound as well as the population dynamics (top) and the corresponding Rabi frequency (bottom) for this model. Using both numerical simulations and an analytical model based on a gauge transformation~\footnote{See Supplemental Material for details.}, we find excellent agreement between theory and simulations~\footnote{This simulation is conducted using the momentum eigenbasis and the fast SDK limit where secular motion is frozen.}. The constant-envelope CW SDK suppresses the backward (counter-rotating) term in the resonant regime, yielding an infidelity of~$1-\mathcal{F}=1.9\times10^{-5}$. This establishes a well-controlled resonant limit that serves as the baseline for exploring optimized pulse shaping and micromotion effects in subsequent sections. 

We next optimize the pulse shape during a single SDK in the CW regime to further suppress residual errors and counter-rotating contributions. In Fig.~\ref{fig: 2}(b), we consider a sine-shaped Rabi envelope of the same~$\tau = 5~\mathrm{ns}$ duration, which provides a smooth turn-on and turn-off. Using the same simulation tool as in Fig.~\ref{fig: 2}(a), we find that the sine envelope minimizes spectral leakage outside the resonant band and strongly reduces coupling to the backward-kick component. With an optimized Raman beat frequency of $\Delta\omega = (1 + 5\times10^{-5}) \omegaa$ and a total pulse area of~$\theta = \pi$, the infidelity of a single CW SDK reaches~$1.4\times10^{-9}$, well below the spontaneous-emission limit of~${}^{133}\mathrm{Ba}^{+}$ driven by 532 nm Raman beams which we calculate to be~$\sim 10^{-7}$~\cite{boguslawski_raman_2023}. This demonstrates that a properly shaped nanosecond-scale pulse can realize near perfect SDKs even without invoking ultrashort pulsed operation~\cite{campbell_ultrafast_2010}.

For comparison with previous work, we also analyze the pulsed scheme originally proposed in the literature~\cite{duan_scaling_2004, campbell_ultrafast_2010, mizrahi_ultrafast_2013_paper}. Guided by the intuition gained from the optimized CW scheme, we construct a sequence of ten~$10~\mathrm{ps}$ pulses with initial amplitudes sampled from a sine envelope and spaced within a total duration of~$5~\mathrm{ns}$. We then optimize the pulse amplitudes, the Raman beat frequency, and the repetition rate~$\omegarep$. The results of the optimization are shown in Fig.~\ref{fig: 2}(c). This pulsed configuration can be viewed as a discrete Trotterization of the continuous interaction. The errors of the Trotterization have been compensated by adjusting the peak amplitudes of the Rabi frequency and optimizing the Raman beat frequency and the repetition rate to~$\Delta\omega = 0.027\omegaa$ and~$\omega_\mathrm{rep} = 2\pi\times1.9~\mathrm{GHz}$, yielding an infidelity of~$3.0\times10^{-9}$ for a single pulsed SDK~\footnote{The final populations of the~$\ket{0, 0}$,~$\ket{1, -2\iu\eta}$, and~$\ket{0, 4\iu\eta}$ states are below~$10^{-9}$. These are recorded as~$10^{-9}$ in Fig.~\ref{fig: 2} for illustration purposes.}. The agreement between the optimized CW and pulsed schemes highlights that the CW model captures the essential dynamics of fast SDKs while requiring nearly~50 times lower peak Rabi frequency.

\begin{figure}[H]
    \centering
    \includegraphics[width=0.95\linewidth]{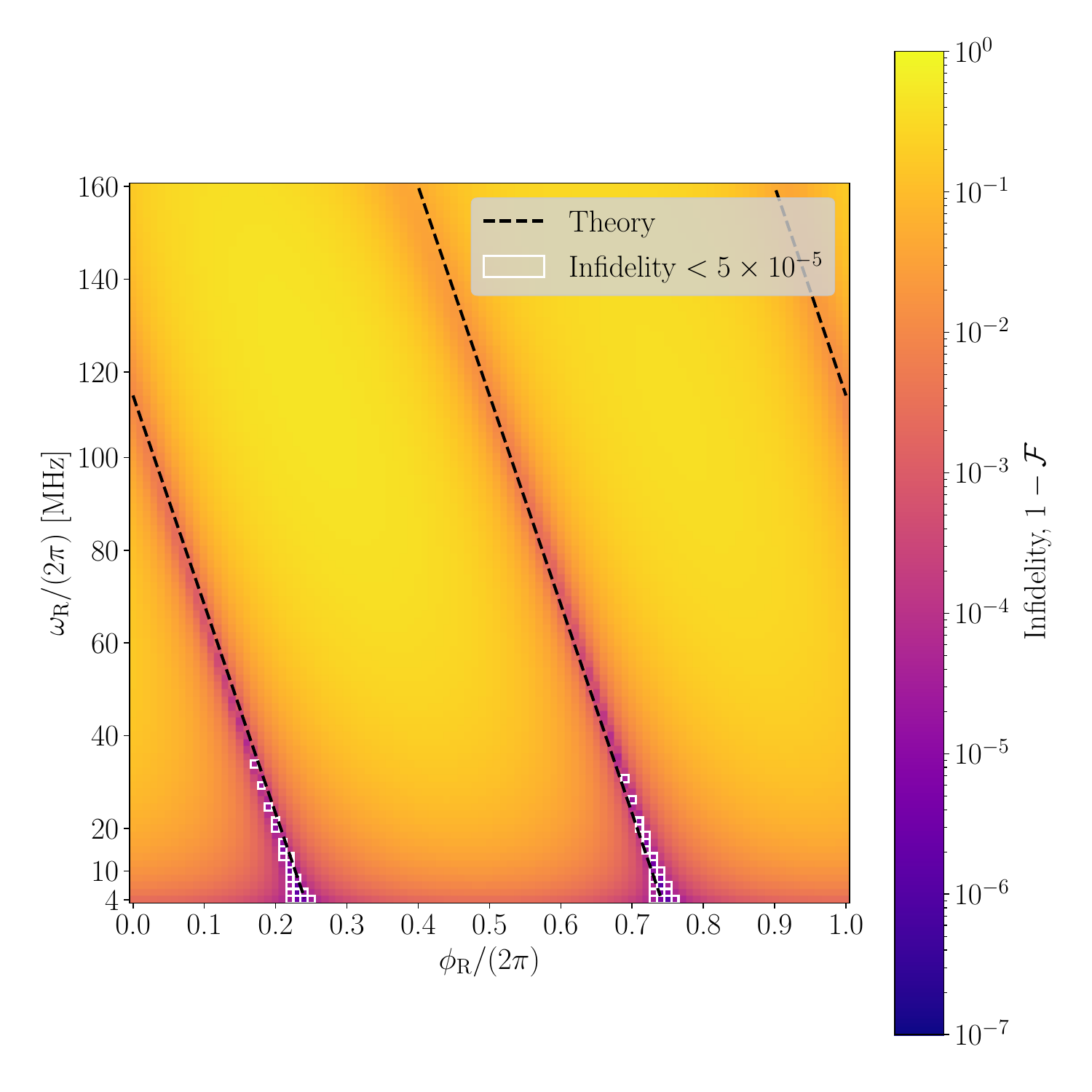}
    \caption{Infidelity landscape of sine-shaped CW SDKs as a function of the RF frequency~$\omegaR$ and RF phase~$\phiR$ using a full quantum simulation that incorporates both secular motion and micromotion effects. The regions where micromotion effects are suppressed agree with the analytical prediction under the fast SDK approximation (black dotted line). White boxed regions indicate parameter regimes yielding SDK infidelity below~$5 \times 10^{-5}$ even in the presence of micromotion. The parameters used are identical to those in Fig.~\ref{fig: 2}(b), with additional realistic Mathieu parameters in the $z$ direction,~$a_z = 0$,~$q_z = 0.15$, and the Lamb-Dicke parameter~$\eta = 0.1$.}
    \label{fig: 3}
\end{figure}

Having established the performance of CW SDKs in the absence of micromotion, we now include the micromotion term in our analysis. In realistic Paul traps, the ion undergoes driven motion at the RF frequency~$\omegaR$, which modulates the phase of the evolution and can significantly affect the SDK fidelity when the SDK duration approaches the micromotion timescale. To tackle this issue, we work in the fast SDK regime given by Eq.~\eqref{eq: fast_SDK_limit} where we ignore the secular motion. In this limit, the micromotion term in Eq.~\eqref{eq: H_IP} commutes with itself at different times. Therefore, by setting the time evolution of the pure micromotion term to identity, we are able to derive an analytical expression for vanishing micromotion effects~\footnote{See Supplemental Material for details.}, i.e.,
\begin{align}
\label{eq: condition_micromotion_null}
    \omegaR \qty(t_0 + \frac{\tau}{2}) + \phiR = \qty(2n + 1)\frac{\pi}{2}, \quad n \in \integers,
\end{align}
where $t_0$ is the initial time of the SDK. Eq.~\eqref{eq: condition_micromotion_null} suggests that by choosing the correct initial phase and frequency of the RF drive, micromotion effects are trivial in the fast SDK limit. To capture the exact dynamics beyond the fast SDK approximation, we numerically simulate the full Hamiltonian, including both secular motion and micromotion effects.
Fig.~\ref{fig: 3} shows the infidelity landscape of a sine-shaped CW SDK as a function of the RF drive frequency~$\omegaR$ and RF phase~$\phiR$ calculated using the full simulation, revealing periodic variations in fidelity that arise from the phase of the driven motion. The black dotted line corresponds to the analytical condition given in Eq.~\eqref{eq: condition_micromotion_null} for micromotion phase matching, where the effective RF phase modulation averages to zero. In this regime, the micromotion-induced sidebands are coherently canceled, and the SDK fidelity is maximized. The white boxed regions mark combinations of~$\omegaR$ and~$\phiR$ yielding infidelity below~$5\times10^{-5}$. These results confirm that micromotion can be effectively suppressed when the SDK is synchronized with the RF drive, establishing the operational regime for high-fidelity fast SDKs in Paul traps. Such RF phase matching has been demonstrated in related contexts and is experimentally feasible~\cite{johnson_active_2016, ratcliffe_micromotion-enhanced_2020}.

\begin{figure}[H]
    \centering
    \includegraphics[width=1.0\linewidth]{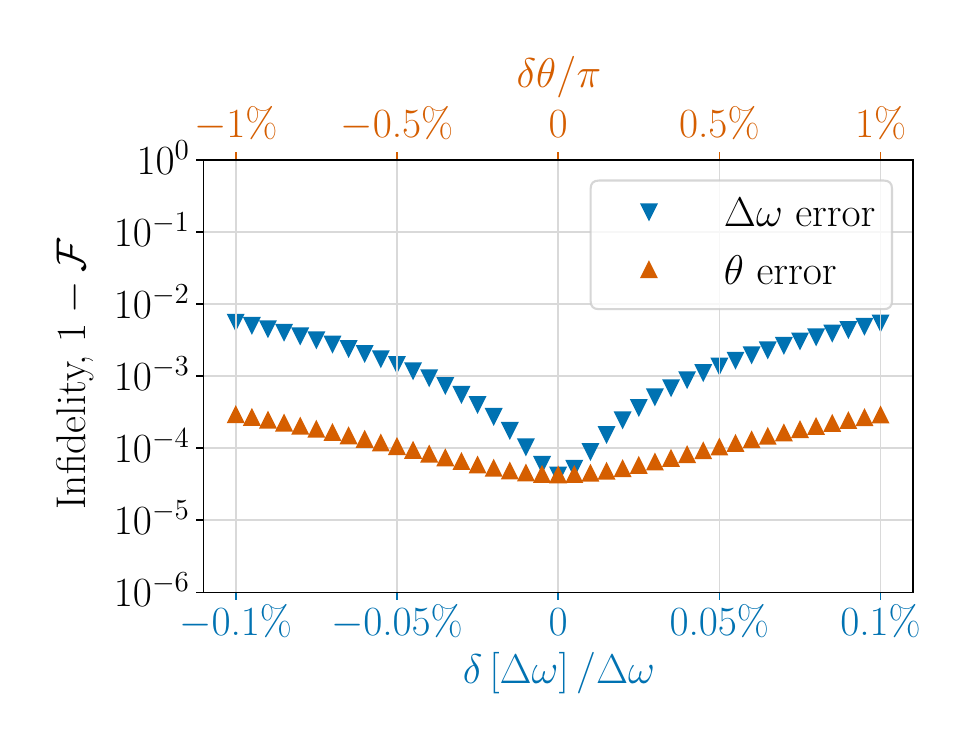}
    \caption{Infidelity of a sine-shaped CW SDK as a function of the errors in pulse area~$\delta\theta$ and Raman beat frequency~$\delta\qty[\Delta\omega]$, respectively. We consider up to~$1\%$ error in pulse area from~$\theta = \pi$ and up to~$0.1\%$ error in Raman beat from~$\Delta\omega = \qty(1 + 5 \times 10^{-5}) \omegaa$. The infidelity remains below~$10^{-3}$ across the pulse area sweep and below~$10^{-2}$ across the Raman beat sweep. These calculations assume one set of optimal RF parameters identified in Fig.~\ref{fig: 3} with~$\omegaR = 2\pi \times 33.64~\mathrm{MHz}$ and~$\phiR = 2\pi \times 0.17$.}
    \label{fig: 4}
\end{figure}

To assess the practical viability of our protocol, we analyze its robustness to common sources of experimental noise. Fig.~\ref{fig: 4} shows infidelity as a function of pulse-area error~$\delta\theta$ and Raman beat frequency error~$\delta\qty[\Delta\omega]$. We consider up to~$1\%$ deviation in pulse area from~$\theta=\pi$ and up to~$0.1\%$ deviation in Ramen beat frequency from~$\Delta\omega = (1 + 5\times 10^{-5})\omegaa$. The SDK remains highly robust, with infidelity below~$10^{-3}$ for pulse-area variations and below~$10^{-2}$ for frequency errors. These calculations use the optimal RF parameters identified in Fig.~\ref{fig: 3}, with $\omegaR=2\pi\times33.64~\mathrm{MHz}$ and~$\phiR=0.17\times2\pi$. The results demonstrate that the optimized CW SDK tolerates realistic control fluctuations while maintaining gate fidelities well below typical error thresholds. 

In conclusion, we have presented a comprehensive framework for realizing high-fidelity, nanosecond-scale SDKs in a CW scheme. Our protocol incorporates the critical, often-neglected effects of intrinsic micromotion and achieves theoretical infidelities below~$10^{-5}$ under experimentally realistic parameters. While SDKs form the foundation of fast two-qubit gates, they also serve as key building blocks for quantum simulation, precision metrology, and the generation of nonclassical motional states~\cite{mizrahi_ultrafast_2013, johnson_sensing_2015}. The results reported here establish a clear and practical pathway toward significantly faster and more precise quantum operations in trapped-ion systems.

Future work could integrate SDK optimization directly into a global optimization of full two-qubit gate operations, treating the SDKs and motional dynamics within a unified cost function. Additionally, extending this analysis to include systematic errors such as laser-pointing deviations or excess micromotion would further refine the achievable gate fidelities and broaden the applicability of our method to large-scale trapped-ion quantum processors. Our framework also enables the design of large-momentum-transfer Raman SDK sequences, where successive kicks add coherently to produce enhanced momentum separation, in the same spirit as large-momentum-transfer Bragg pulses developed for atom-interferometric beam splitters~\cite{muller_atom_2008, mazzoni_large-momentum-transfer_2015, giltner_atom_1995, kasevich_atomic_1991}.

\section{Acknowledgments} 
The authors would like to thank J. J. Hope, Z. Mehdi, I. Savill-Brown, C. Sagaseta, E. Torrontegui, J. J. Garc\'ia-Ripoll, I. D. Moore, and J. Amini for useful discussions.

\bibliographystyle{apsrev4-2}
\bibliography{main2}

\end{document}


\title{Supplemental Material: High-Fidelity Raman Spin-Dependent Kicks in the Presence of Micromotion}

\author{Haonan Liu}
\email{haonan.liu@ionq.co}
\affiliation{IonQ, Inc., College Park, MD, USA}
\author{Varun D. Vaidya}
\affiliation{IonQ, Inc., College Park, MD, USA}
\author{Monica Gutierrez Galan}
\affiliation{IonQ, Inc., College Park, MD, USA}
\author{Alexander K. Ratcliffe}
\affiliation{IonQ, Inc., College Park, MD, USA}%
\author{Amrit Poudel}
\affiliation{IonQ, Inc., College Park, MD, USA}
\author{C. Ricardo Viteri}
\affiliation{IonQ, Inc., College Park, MD, USA}

\date{\today}
\maketitle

\section{\label{appx: hamiltonian}Hamiltonian}
In this section, we derive the Hamiltonians given in Eq.~(1)--(2) used in the main text from the first principles.
\subsection{\label{appx: su3}$\Lambda$-type Raman three-level system interacting with a single field}
For simplicity, we start from deriving the Hamiltonian of a $\Lambda$-type Raman three-level system interacting with a single field. We only include the internal Hilbert space of the system for now.
\subsubsection{Three-level Hamiltonian}
Consider a three-level system with levels $\qty{\ket{0}, \ket{1}, \ket{2}}$. Without loss of generality, we refer to $\ket{0}$ as the ground state,  $\ket{1}$ as the metastable state, and $\ket{2}$ as the excited state. Here, $\qty{\ket{0}, \ket{1}}$ are the qubit states. The Hamiltonian of such a system is given by
\begin{align}
\label{eq: H_a_1}
    H_\mathrm{a} = \sum_{j=0}^{2}\omega_j\ketbra{j}{j} = \mqty*(\dmat{\omega_2, \omega_1, \omega_0}),
\end{align}
where $\omega_j$ is the corresponding frequency of the $j$th energy level. We have set $\hbar = 1$ for convenience.

For simplicity, we start by studying the interaction between the three-level system and a single classical field, given by
\begin{align}
\label{eq: field_1}
    \vb{E} \qty(\vb{r}, t) = \Re\qty[E(t) e^{\iu \qty(\vb{k}\cdot\vb{r} - \omega t)} \bm{\epsilon}] = \frac{1}{2} \qty[E(t) e^{\iu \qty(\vb{k}\cdot\vb{r} - \omega t)} \bm{\epsilon} + E^\ast(t)e^{-\iu \qty(\vb{k}\cdot\vb{r} - \omega t)} \bm{\epsilon}^\ast].
\end{align}
Here, the positive frequency $\omega$ is a characteristic frequency of the field (laser center frequency for example). The polarization vector $\bm{\epsilon}$ can be complex. The field envelope~$E(t)$ is a complex-valued pulsed envelope, which can be understood as a time-dependent real non-oscillating amplitude~$\mathcal{E}(t)$ multiplied by a phase factor that can be time-dependent, i.e.,
\begin{align}
\label{eq: field_2}
    E(t) = \mathcal{E}(t) e^{-\iu \phi(t)}.
\end{align}
The Hamiltonian of the interaction between an atom at position~$\vb{r}$ and a classical field $\vb{E} \qty(\vb{r}, t)$ is given by the~$\vb{d}\cdot\vb{E}$ form,
\begin{align}
\label{eq: H_I_d_dot_e}
    H_\mathrm{I}(t) & = - \sum_{j,j'=0}^{2} \ketbra{j}{j'}\vb{d}_{jj'} \cdot \vb{E}(t) \\
\label{eq: H_I_middle}
    & = - \frac{1}{2} \sum_{j,j'=0}^{2} \ketbra{j}{j'} \qty[e^{-\iu \omega t} E(t) e^{\iu \vb{k}\cdot\vb{r}} \vb{d}_{jj'}\cdot\bm{\epsilon} + e^{\iu \omega t} E^\ast(t) e^{-\iu \vb{k}\cdot\vb{r}}\vb{d}_{jj'}\cdot\bm{\epsilon}^\ast ]
\end{align}
where $\vb{d}_{jj'}$'s are the complex dipole matrix elements~\footnote{We note that in Eq.~\eqref{eq: H_I_middle}, the dot product is not defined by the Hermitian form~$\vb{u}\cdot\vb{v} = \sum u_j v_j^\ast$. Rather, it follows the definition of the dot product of two real vectors since the field~$\vb{E}$ is real anyway.}

We now define the single photon Rabi frequencies as
\begin{align}
\label{eq: gjj'}
    g_{jj'}(t) &= - E(t) e^{\iu \vb{k}\cdot\vb{r}}\vb{d}_{jj'} \cdot \bm{\epsilon}.
\end{align}
The dipole matrix is Hermitian, giving the following relationship
\begin{align}
\label{eq: djj}
    \vb{d}_{jj'} = \vb{d}^\ast_{j'j},
\end{align}
which yields
\begin{align}
\label{eq: gj'j_ast}
    g_{j'j}^\ast(t) = - E^\ast(t) e^{-\iu \vb{k}\cdot\vb{r}}\vb{d}_{j'j}^\ast \cdot \bm{\epsilon}^\ast = -E^\ast(t) e^{-\iu \vb{k}\cdot\vb{r}}\vb{d}_{jj'} \cdot \bm{\epsilon}^\ast.
\end{align}
Substituting Eqs.~\eqref{eq: gjj'} and~\eqref{eq: gj'j_ast} into Eq.~\eqref{eq: H_I_middle}, we have
\begin{align}
\label{eq: H_I_1}
    H_\mathrm{I}(t) = \frac{1}{2}\sum_{j,j'=0}^{2} \ketbra{j}{j'} \qty[g_{jj'}(t) e^{-\iu\omega t} +  g_{j'j}^\ast(t) e^{\iu\omega t}].
\end{align}

We now assume that the three-level system is a $\Lambda$-type Raman system such that the field can only drive the $\ket{0}\leftrightarrow\ket{2}$ and $\ket{1}\leftrightarrow\ket{2}$ transitions. Then the interaction Hamiltonian in~Eq.~\eqref{eq: H_I_1} can be written as the following $3\times 3$ Hermitian matrix
\begin{align}
\label{eq: H_I_3}
    H_\mathrm{I} = \frac{1}{2}\mqty*(& g_{12} e^{-\iu \omega t} +  g_{21}^\ast e^{\iu \omega t} & g_{02} e^{-\iu \omega t} +  g_{20}^\ast e^{\iu \omega t}\\ g_{21} e^{-\iu \omega t} +  g_{12}^\ast e^{\iu \omega t} & & \\g_{20} e^{-\iu \omega t} +  g_{02}^\ast e^{\iu \omega t} & &).
\end{align}
The total Hamiltonian is thus
\begin{align}
    H = H_\mathrm{a} + H_\mathrm{I}.
\end{align}
\subsubsection{The $\su{3}$ Lie algebra}
Equations~\eqref{eq: H_a_1} and~\eqref{eq: H_I_3} strongly suggest symmetry. Indeed the total Hamiltonian can be expressed by the~$\su{3}$ generators in their defining representation. By convention, the $\su{3}$ generators are usually given by the Gell-Mann matrices, i.e.,
\begin{align}
\label{eq: Gell-Mann_1}
    \lambda_1 &= \mqty*(\dmat{\pmat{1}, 0}), \quad \lambda_2 = \mqty*(\dmat{\pmat{2}, 0}), \quad \lambda_3 = \mqty*(\dmat{\pmat{3}, 0}), \\
    \lambda_4 &= \mqty*(\admat{1,0,1}), \quad \lambda_5 = \mqty*(\admat{-\iu,0,\iu}),\\
\label{eq: Gell-Mann_3}
    \lambda_6 &= \mqty*(\dmat{0, \pmat{1}}), \quad \lambda_7 = \mqty*(\dmat{0, \pmat{2}}), \quad \lambda_8 = \frac{1}{\sqrt{3}}\mqty*(\dmat{1,1,-2}).
\end{align}
We can identify the $\su{2}$ subalgebras of the $\su{3}$ algebra by reorganizing the generators into a new set of basis operators. These subalgebras are manifest if we define the following operators
\begin{align}
    \threechoices{T_+ = \frac{\lambda_1 + \iu \lambda_2}{2} = \mqty*(\dmat{0&1\\0&0,0})}{T_- = \frac{\lambda_1 - \iu \lambda_2}{2} = \mqty*(\dmat{0&0\\1&0,0})}{T_3 = \lambda_3 = \mqty*(\dmat{\pmat{3}, 0})},
    \threechoices{U_+ = \frac{\lambda_6 + \iu \lambda_7}{2} = \mqty*(\dmat{0, 0&1\\0&0})}{U_- = \frac{\lambda_6 - \iu \lambda_7}{2} = \mqty*(\dmat{0,0&0\\1&0})}{U_3 = \frac{-\lambda_3 + \sqrt{3}\lambda_8}{2} = \mqty*(\dmat{0, \pmat{3}})},
    \threechoices{V_+ = \frac{\lambda_4 + \iu \lambda_5}{2} = \mqty*(\admat{1,0,0})}{V_- = \frac{\lambda_4 - \iu \lambda_5}{2} = \mqty*(\admat{0,0,1})}{V_3 = \frac{\lambda_3 + \sqrt{3}\lambda_8}{2} = \mqty*(\dmat{1,0,-1})}.
\end{align}
The subalgebra with generators $\qty{\lambda_1, \lambda_2, T_3}$ is isomorphic to $\su{2}$ (i.e., $\qty{\sigma_x, \sigma_y, \sigma_z}$), with raising and lowering operators $T_\pm$ corresponding to $\sigma_\pm$. So on for the $U$ and $V$ subalgebras.

Each of the $\su{2}$ subalgebras corresponds to a two-level transition, i.e., $T$ for the~$\ket{1}\leftrightarrow\ket{2}$ transition, $U$ for the~$\ket{0}\leftrightarrow\ket{1}$ transition, and $V$ for the~$\ket{0}\leftrightarrow\ket{2}$ transition. Since we only have the~$\ket{1}\leftrightarrow\ket{2}$ and~$\ket{0}\leftrightarrow\ket{2}$ transitions in the Hamiltonian, it is expected that our Hamiltonian only have the $T$ and $V$ components in the interaction part. Indeed, we can rewrite Eq.~\eqref{eq: H_I_3} as
\begin{align}
\label{eq: H_I_2}
    H_\mathrm{I} = \frac{1}{2}\qty(g_{12}T_+ + g_{21} T_- + g_{02}V_+ + g_{20} V_-) e^{-\iu \omega t} + \frac{1}{2}\qty(g_{21}^\ast T_+ + g_{12}^\ast T_- + g_{20}^\ast V_+ + g_{02}^\ast V_-) e^{\iu \omega t}.
\end{align}
The internal Hamiltonian in Eq.~\eqref{eq: H_a_1} can also be rewritten as
\begin{align}
\label{eq: H_a_2}
    H_\mathrm{a} = \frac{\omega_2 - \omega_1}{3} T_3 + \frac{\omega_1 - \omega_0}{3} U_3 + \frac{\omega_2 - \omega_0}{3} V_3 + \frac{\omega_2+\omega_1+\omega_0}{3}\identity_3.
\end{align}
The identity matrix needs to be included because the $\su{3}$ generators only span the space of traceless Hermitian matrices. This way, the total Hamiltonian of the Raman system has been decomposed into $\su{3}$ generators plus the identity matrix.
\subsubsection{Interaction picture}
It is convenient to study the total Hamiltonian in the interaction picture (IP) defined by $H_\mathrm{a}$, where the new Hamiltonian is given by~$H = U^\dagger H_\mathrm{I} U$, where~$U = \exp\qty(- \iu H_\mathrm{a} t)$. This conjugation on $H_\mathrm{I}$ can be analytically solved. Specifically, we use the following result~\footnote{This can be seen from several viewpoints. One may evaluate it directly using the Baker–Campbell–Hausdorff expansion. A more systematic perspective is to note that, in the adjoint representation, $x$ is the eigenvalue of $A$ acting on the eigenvector $B$; consequently, $e^{A}$ acts with eigenvalue $e^{x}$. In many physics texts, the map $[A,\cdot]$ is introduced as a ``superoperator'' acting on $B$, which in this context is precisely the adjoint action of $A$.}.
\begin{theorem}
\label{thm: ajoint}
    For $A, B$ in some Lie algebra, if $\comm{A}{B} = xB$, then $e^A B e^{-A} = e^x B$.
\end{theorem}
Applying Thm.~\eqref{thm: ajoint} to Eqs.~\eqref{eq: H_I_2}--\eqref{eq: H_a_2}, we have
\begin{align}
    [\iu t H_\mathrm{a}, T_\pm] &= \pm \iu (\omega_2 - \omega_1)t T_\pm,\\
    [\iu t H_\mathrm{a}, V_\pm] &= \pm \iu (\omega_2 - \omega_0)t V_\pm,
\end{align}
and immediately
\begin{align}
    H =& e^{\iu t H_\mathrm{a}} H_\mathrm{I} e^{-\iu t H_\mathrm{a}} \\
    =& \frac{1}{2}e^{\iu t H_\mathrm{a}}\qty[\qty(g_{12}T_+ + g_{21} T_- + g_{02}V_+ + g_{20} V_-) e^{-\iu \omega t} + \qty(g_{21}^\ast T_+ + g_{12}^\ast T_- + g_{20}^\ast V_+ + g_{02}^\ast V_-)e^{\iu \omega t}] e^{-\iu t H_\mathrm{a}} \\
    =&\quad\, \frac{1}{2}\qty[g_{12}T_+ e^{\iu(\omega_2-\omega_1)t} + g_{21} T_- e^{-\iu(\omega_2-\omega_1)t}+ g_{02} V_+ e^{\iu(\omega_2-\omega_0)t} + g_{20} V_- e^{-\iu(\omega_2-\omega_0)t}] e^{-\iu \omega t} \notag\\
\label{eq: H_IP}
    &+ \frac{1}{2}\qty[g_{21}^\ast T_+ e^{\iu(\omega_2-\omega_1)t} + g_{12}^\ast T_- e^{-\iu(\omega_2-\omega_1)t} + g_{20}^\ast V_+ e^{\iu(\omega_2-\omega_0)t}+ g_{02}^\ast V_- e^{-\iu(\omega_2-\omega_0)t}] e^{\iu \omega t}.
\end{align}
Therefore the effect of entering the IP is introducing a phase factor to the raising and lowering operators, i.e.,
\begin{align}
    T_\pm \rightarrow T_\pm e^{\pm \iu\qty(\omega_2 - \omega_1)}, 
    \quad
    V_\pm \rightarrow V_\pm e^{\pm \iu\qty(\omega_2 - \omega_0)}.
\end{align}
\subsubsection{Rotating wave approximation}
The Hamiltonian in Eq.~\eqref{eq: H_IP} is composed of terms oscillating with multiple frequencies. For convenience, we define two useful frequencies:
\begin{enumerate}
    \item The qubit transition frequency between levels $\ket{1}\leftrightarrow\ket{0}$
    \begin{align}
    \label{eq: omegaa}
        \omegaa = \omega_1 - \omega_0.
    \end{align}

    \item The laser detuning from the $\ket{2}\leftrightarrow\ket{0}$ transition
    \begin{align}
    \label{eq: detuning}
        \Delta = \omega - (\omega_2 - \omega_0).
    \end{align}
\end{enumerate}
These frequencies have been introduced in Fig.~1(b) of the main text. Then naturally the laser detuning from the~$\ket{2}\leftrightarrow\ket{1}$ transition is given by~$\Delta + \omegaa$. In a Raman system, we expect the following hierarchy of frequencies:
\begin{approximation}[Rotating wave approximation]
\label{approx: RWA}
    \begin{align}
        \omegaa \lesssim \abs{\frac{\dot{\mathcal{E}}}{\mathcal{E}}}, \abs{\dot{\phi} (t)} \ll \Delta \ll \omega + \omega_2.
    \end{align}
\end{approximation}
Here, $\dot{\mathcal{E}}$ accounts for the non-oscillating change in the field amplitude, and $\phi(t)$ accounts for the oscillating phase of the field. Given the frequency hierarchy in Approx.~\eqref{approx: RWA}, we can ignore the terms with high frequencies $\pm\qty(\omega+\omega_2)$ in $H$, resulting in
\begin{align}
\label{eq: H_su3}
   H = \frac{1}{2}\qty[
   g_{12} T_+ e^{-\iu(\Delta + \omegaa)t} + 
   g_{02} V_+ e^{-\iu\Delta t}]
   + \frac{1}{2}\qty[
   g_{12}^\ast T_- e^{\iu(\Delta + \omegaa)t} + 
   g_{02}^\ast V_- e^{\iu\Delta t}].
\end{align}

This is the rotating-wave approximation (RWA). Eq.~\eqref{eq: H_su3} gives the final form of the~$\su{3}$ Hamiltonian in the interaction picture.
\subsubsection{Adiabatic elimination}
The Hamiltonian in Eq.~\eqref{eq: H_su3} can be further reduced. Remember our ultimate goal is to use levels~0 and~1 as qubit levels. Therefore it would be nice to have an effective Hamiltonian that acts only on the~$\qty{\ket{0}, \ket{1}}$ submanifold. This entails a twofold interpretation:
\begin{enumerate}
    \item Mathematically this means the effective Hamiltonian should be block diagonal and in the subalgebra spanned by $\qty{U_\pm, U_3}$. We are going from~$\su{3}$ to one of its~$\su{2}$ subalgebra. 
    \item Physically this means that the excited state $\ket{2}$ has a far detuned transition to the qubit states from the field and therefore is weakly populated. Luckily, this is indeed the case as we will see.
\end{enumerate}

Consider the Schr\"odinger equation~\footnote{One can also use the Schrieffer–Wolff Transformation for this purpose. Here we show one approach.}
\begin{align}
\label{eq: schrodinger}
    \iu \dot{\ket{\psi}} = H \ket{\psi},
\end{align}
where $\ket{\psi} = \sum_j c_j \ket{j}$. Left multiplication by $\bra{j}$ yields
\begin{align}
\label{eq: c0}
    \iu \dot{c_0} &= \frac{g_{02}^\ast}{2} e^{\iu \Delta t} c_2,\\
\label{eq: c1}
    \iu \dot{c_1} &= \frac{g_{12}^\ast}{2} e^{\iu \qty(\Delta+\omegaa) t} c_2, \\
\label{eq: c2}
    \iu \dot{c_2} &= \frac{g_{12}(t)}{2} e^{-\iu(\Delta + \omegaa)t} c_1 + \frac{g_{02}(t)}{2} e^{-\iu\Delta t} c_0.
\end{align}
Formally solving for $c_2(t)$ yields
\begin{align}
\label{eq: AE}
    c_2(t) = - \iu \int_0^t \dd{u} \frac{g_{12}(u)}{2} c_1(u) e^{-\iu(\Delta + \omegaa)u} - \iu \int_0^t \dd{u} \frac{g_{02}(u)}{2} c_0(u) e^{-\iu\Delta u},
\end{align}
where $c_2(0)$ has been chosen to be 0. 

We now make the approximation of adiabatic elimination by considering the following frequency hierarchy:
\begin{approximation}[Adiabatic elimination, 2$^\mathrm{nd}$ rotating wave approximation]
\label{approx: AE}
    \begin{align}
\label{eq: approx: AE}
        \abs{\dot{c_1}(t)}, \abs{\dot{c_2}(t)}, \abs{\frac{\dot{\mathcal{E}}}{\mathcal{E}}}, \abs{\dot{\phi}(t)} \ll \Delta,
    \end{align}
\end{approximation}
The adiabatic elimination approximation argues that the detunings of the field from both of the~$\ket{2}\leftrightarrow\ket{0}$ and~$\ket{2}\leftrightarrow\ket{1}$ transitions are so large that all the other time-varying factors inside the integrals in Eq.~\eqref{eq: AE} can be taken as a constant with $u=t$. In other words, $c_2$ can be directly solved as a function of~$c_1$ and~$c_0$. 

Comparing Approx.~\eqref{approx: AE} to Approx.~\eqref{approx: RWA} we see that the two approximations have overlaps in that the detuning~$\Delta$ needs to dominate the field changing rate. However, adiabatic elimination further requires~$\Delta$ to dominate~$\abs{\dot{c_1}(t)}$ and~$\abs{\dot{c_2}(t)}$, which is more or less heuristic. Following the procedure of adiabatic elimination, we thus have
\begin{align}
\label{eq: c2_integral_1}
    c_2(t) &\approx -\iu \frac{g_{12}(t)}{2} c_1(t) \int_0^t \dd{u} e^{-\iu \qty(\Delta + \omegaa) u} -\iu \frac{g_{02}(t)}{2} c_0(t) \int_0^t \dd{u} e^{-\iu \Delta u}\\
\label{eq: c2_sol_1}
    &= \frac{e^{-\iu \qty(\Delta + \omegaa) t} - 1}{\Delta + \omegaa} \frac{g_{12}(t)}{2} c_1(t) + \frac{e^{-\iu \Delta t} - 1}{\Delta} \frac{g_{02}(t)}{2} c_0(t).
\end{align}
Substituting Eq.\eqref{eq: c2_sol_1} into Eqs.~\eqref{eq: c0}--\eqref{eq: c1} yields
\begin{align}
\label{eq: c1_sol_AE}
    \iu \dot{c_1} &= \frac{1 - e^{\iu \qty(\Delta + \omegaa) t}}{\Delta + \omegaa} \frac{\abs{g_{12}}^2}{4} c_1 + \frac{1 - e^{\iu \Delta t}}{\Delta} e^{\iu \omegaa t} \frac{g_{12} g_{02}^\ast}{4} c_0, \\
\label{eq: c0_sol_AE}
    \iu \dot{c_0} &= \frac{1 - e^{-\iu \qty(\Delta + \omegaa) t}}{\Delta + \omegaa} e^{- \iu \omegaa t} \frac{g_{02}g_{12}^\ast}{4} c_1 + \frac{1 - e^{\iu \Delta t}}{\Delta} \frac{\abs{g_{02}}^2}{4} c_0.
\end{align}
\subsubsection{Effective two-level Hamiltonian}
The results from the adiabatic elimination, given in Eqs.~\eqref{eq: c1_sol_AE}--\eqref{eq: c0_sol_AE}, correspond to a non-Hermitian Hamiltonian. Although we now have an effective two-level system, the probability for an atom to leak from the qubit manifold into the excited state $\ket{2}$ still scales with $1/\Delta$ and is thus nonzero. In order to work in the two-level system, we must engineer our Hamiltonian with given approximations to make it Hermitian. 

One approach is as follows. We first replace all the $\Delta + \omegaa$ terms by $\Delta$, which follows from the RWA in Approx.~\eqref{approx: RWA}, yielding
\begin{align}
    \iu \dot{c_1} &= \frac{1 - e^{\iu \Delta t}}{4\Delta} \qty(\abs{g_{12}}^2 c_1 + e^{\iu \omegaa t} g_{12} g_{02}^\ast c_0), \\
    \iu \dot{c_0} &= \frac{1 - e^{\iu \Delta t}}{4\Delta} \qty(e^{- \iu \omegaa t} g_{02}g_{12}^\ast c_1 + \abs{g_{02}}^2 c_0).
\end{align}
At this point we make another RWA, which recognizes that $\Delta$ is much larger than the other frequencies (essentially Approx.~\eqref{approx: AE} again), so that the $e^{\iu \Delta t}$ terms can be averaged out. We end up with
\begin{align}
\label{eq: Heff_0}
    H_\mathrm{eff} = \frac{1}{4\Delta}
    \mqty*(\abs{g_{12}}^2 & e^{\iu \omegaa t} g_{12}g_{02}^\ast\\
    e^{- \iu \omegaa t} g_{02}g_{12}^\ast & \abs{g_{02}}^2).
\end{align}
Defining the effective Rabi frequency (or two-photon Rabi frequency) and the two-photon light shifts as
\begin{align}
\label{eq: Rabi}
    \Omega(t) = \frac{g_{12}g_{02}^\ast}{2\Delta},\quad 
    \delta_1(t) = \frac{\abs{g_{12}}^2}{4\Delta},\quad
    \delta_0(t) = \frac{\abs{g_{02}}^2}{4\Delta},
\end{align}
we have
\begin{align}
\label{eq: Heff_1}
    H_\mathrm{eff} = \frac{\delta_1+\delta_0}{2}\identity + e^{\iu \omegaa t} \frac{\Omega}{2} \sigma_+ + e^{- \iu \omegaa t} \frac{\Omega^\ast}{2} \sigma_- + \frac{\delta_1 - \delta_0}{2} \sigma_z.
\end{align}
The $\sigma_\pm$ and $\sigma_z$ operators in $\su{2}$ are none other than the $U_\pm$ and $U_3$ operators in $\su{3}$, which drive the transition~$\ket{1}\leftrightarrow\ket{0}$.

By Thm.~\eqref{thm: ajoint}, Eq.~\eqref{eq: Heff_1} can be taken as the effective Hamiltonian of a two-level system in the IP defined by $H_\mathrm{a} = \displaystyle\frac{\omegaa}{2} \sigma_z$. In the effective Schr\"odinger picture (SP), we thus have
\begin{align}
\label{eq: Heff_1_SP}
    H_\mathrm{eff} = \frac{\delta_1+\delta_0}{2}\identity + \frac{\Omega}{2} \sigma_+ + \frac{\Omega^\ast}{2} \sigma_- + \frac{\omegaa + \delta_1 - \delta_0}{2} \sigma_z.
\end{align}

The first term on the right hand side (RHS) of Eq.~\eqref{eq: Heff_1_SP} corresponds to a time-dependent global phase on the state, given by~$e^{\iu\frac{\delta_0(t) + \delta_1(t)}{2}t}$. We will always work in this IP and ignore the first term from now on. The second term on the RHS corresponds to the interaction and is the key to our study. The third term corresponds to the internal free evolution plus a differential light shift $\delta_1-\delta_0$. We now make the approximation
\begin{approximation}[small differential light shift]
    \begin{align}
        \delta_1(t) - \delta_0(t) \ll \omegaa, \Omega(t)
    \end{align}
\end{approximation}
\noindent which has been discussed in~\cite{mizrahi_ultrafast_2013}. The final effective two-level Hamiltonian is then
\begin{align}
\label{eq: H_one_beam}
    H_\mathrm{eff}(t) = \frac{\Omega(t)}{2} \sigma_+ + \frac{\Omega^\ast(t)}{2} \sigma_- + \frac{\omegaa}{2} \sigma_z.
\end{align}
\subsection{\label{appx: two_beam}General theory: more levels, more fields}

In this section we derive a general theory for a multi-$\Lambda$ system interacting with multiple fields. The theory is based on Sec.~\ref{appx: su3}.
\subsubsection{Rabi frequency}
Following Sec.~\ref{appx: su3}, a field (labeled by $m$) has the general form
\begin{align}
\label{eq: field_general}
    \vb{E}_m \qty(\vb{r}, t) = \Re\qty[\mathcal{E}_m(t) e^{\iu \qty[\vb{k}_m\cdot\vb{r} - \omega t - \phi_m(t)]} \bm{\epsilon}_m],
\end{align}
where~$\mathcal{E}_m(t)$ is the real amplitude envelope,~$\omega$ is the characteristic frequency of the fields,~$\vb{k}_m$ is the wavevector,~$\phi_m(t)$ is a time-dependent phase factor, and~$\bm{\epsilon}_m$ is the normalized polarization vector. Notice that here the frequency $\omega$ does not need to be the center frequency of all the fields. Rather, the difference in center frequency of the $m^{\mathrm{th}}$ field relative to $\omega$ has been absorbed into the phase factor $\phi_m(t)$. For example, for two fields of frequencies $\omega_1$ and $\omega_2$, we can choose $\omega = \omega_1$, $\phi_1(t) = c_1(t)$ and $\phi_2(t) = (\omega_2 - \omega_1)t + c_2(t)$ where $c_1(t)$ and $c_2(t)$ are residue phase terms including the initial phases and noise terms if any.

For each laser field, we have defined the single-photon Rabi frequency in Eq.~\eqref{eq: gjj'} as
\begin{align}
\label{eq: gjj'm}
    \qty(g_{jj'})_m &= - \mathcal{E}_m(t) e^{\iu \vb{k}_m\cdot\vb{r}} e^{ - \iu \phi_m(t)} \vb{d}_{jj'} \cdot \bm{\epsilon}_m.
\end{align}
For multiple laser fields, we can likewise define the collective single-photon Rabi frequency as
\begin{align}
\label{eq: gjj'_general}
    g_{jj'} &= \sum_m \qty(g_{jj'})_m = - \sum_m \mathcal{E}_m(t) e^{\iu \vb{k}_m\cdot\vb{r}} e^{ - \iu \phi_m} \vb{d}_{jj'} \cdot \bm{\epsilon}_m.
\end{align}

We next derive the effective two-level Hamiltonian with multiple fields. Let the polarization vector of the $m^{\mathrm{th}}$ beam be
\begin{align}
\label{eq: polarization_vector_real}
    \bm{\epsilon}_m = \sin{\alpha_m} \cos{\beta_m} e^{\iu\theta_m} \vb{e}_{\sigma_+} +
                    \sin{\alpha_m} \sin{\beta_m} e^{\iu\psi_m} \vb{e}_{\sigma_-} +
                    \cos{\alpha_m} \vb{e}_{\pi},          
\end{align}
where we have chosen the most general polarization vector. Then by selection rules, different polarization components of each field will couple to different energy levels. For each level, we can follow the derivations in Sec.~\ref{appx: su3} and calculate a two-photon Rabi frequency as in Eq.~\eqref{eq: Rabi}. Clearly, in order for the RWA and AE approximations to remain valid, we need to impose for any beam $m$ and level $j$ the following condition
\begin{approximation}[Rotating wave approximation, adiabatic elimination, multiple beams, multiple levels]
\label{approx: RWA&AE_multiple_beam}
    \begin{align}
    \label{eq: RWA&AE_multiple_beam}
        \abs{\dot{\phi}_m(t)} \ll \Delta_j,
    \end{align}
\end{approximation}
\noindent where $\Delta_j$ is the detuning corresponding to the $j$th level
\begin{align}
    \Delta_j = \omega - \qty(\omega_j - \omega_0).
\end{align}
Then we follow~Sec.~\ref{appx: su3} and derive the SP effective two-level Hamiltonian as
\begin{align}
\label{eq: Hamiltonian_no_motion}
    H_\mathrm{eff} = \frac{\Omega(t)}{2} \sigma_+ + \frac{\Omega^\ast(t)}{2} \sigma_- + \frac{\omegaa}{2} \sigma_z,
\end{align}
where $\omegaa$ is the qubit transition frequency and $\Omega(t)$ is the collective two-photon Rabi frequency, which we will refer to as the \textit{Rabi frequency}, given by
\begin{align}
    \Omega(t) &= \sum_{j}\frac{g_{1j}g_{0j}^\ast}{2\Delta_j},
\end{align}
where $g_{1j}$ and $g_{0j}$ are defined in Eq.~\eqref{eq: gjj'_general}. Substituting in Eq.~\eqref{eq: gjj'_general} yields
\begin{align}
\label{eq: omega_multiple}
    \Omega(t) &= \sum_{j=2}^{N-1} \sum_{m,n=1}^M \frac{\mathcal{E}_m(t)\mathcal{E}_{n}(t)}{2\Delta_j} e^{\iu \qty[\qty(\vb{k}_m - \vb{k}_n)\cdot\vb{r} - \qty(\phi_m-\phi_n)]} \qty(\vb{d}_{1j} \cdot \bm{\epsilon}_m) \qty(\vb{d}_{0j} \cdot \bm{\epsilon}_n)^\ast,
\end{align}
where $N$ is the total number of coupled levels (levels 0 and 1 refer to the qubit levels specifically, while levels $j$ to $N-1$ refer to the remaining levels) and $M$ is the total number of fields. The polarization vectors are parametrized using Eq.~\eqref{eq: polarization_vector_real}. 
\subsection{Two collimated beam model}
We are now prepared to analyze a concrete experimental setup, where we consider a case with two collimated laser beams (either co- or counter-propagating) of similar center frequency~$\omega$ along the~$z$ quantization axis. 
\subsubsection{Rabi frequency}
For each beam, the polarization vector is
\begin{align}
\label{eq: polarization_vector_no_pi}
    \bm{\epsilon}_m = \cos{\beta_m} \vb{e}_{\sigma_+} + e^{\iu \psi_m} \sin{\beta_m} \vb{e}_{\sigma_-},   
\end{align}
where $\beta_m$'s are used to parametrize left and right polarizations, and $\psi_m$'s are the relative phases. Notice that a global phase can be manipulated as its counterpart can be absorbed into the initial phase in Eq.~\eqref{eq: field_general}. Due to geometry, neither beam has~$\pi$ polarizations.

We choose the qubit states as the hyperfine states $\twochoices{\ket{1} = \ket{\alpha, F=1, M_F=0}}{\ket{0} = \ket{\alpha, F=0, M_F=0}}$. Notice that in both the cases of~\Yb and~\Ba,~$\alpha$ stands for the ground state $^{2}\mathrm{S}_{1/2}$, but the value is irrelevant in our theoretical derivation. By the dipole selection rules, the qubit states can only be simultaneously coupled to states in the $F=1$ manifolds. Let 
\begin{align}
    \ket{L} &= \ket{\alpha', F=1, M_F = 1},\\
    \ket{R} &= \ket{\alpha', F=1, M_F = -1},
\end{align}
where we have ignored the $M_F = 0$ state due to the fact that the collimated beams do not have $\pi$ polarizations. Then the Wigner-Eckart theorem, we have
\begin{align}
\label{eq: d0_1_L_R}
    \twochoices
    {\vb{d}_{0L} \cdot \bm{\epsilon}_m =  e^{\iu\psi_m} \sin{\beta_m} d_{0}}
    {\vb{d}_{0R} \cdot \bm{\epsilon}_m =  \cos{\beta_m} d_{0}},\quad
    \twochoices
    {\vb{d}_{1L} \cdot \bm{\epsilon}_m =  - e^{\iu\psi_m} \sin{\beta_m} d_{1}}
    {\vb{d}_{1R} \cdot \bm{\epsilon}_m =  \cos{\beta_m} d_{1}},
\end{align}
where $d_0$ and $d_1$ are the refactored reduced dipole matrix elements whose values are irrelevant to our derivation.

We further choose the detunings of the two beams to be identical, i.e.,
\begin{align}
    \Delta_L &= \Delta_R \equiv \Delta.
\end{align}
Then Eq.~\eqref{eq: omega_multiple} becomes
\begin{align}
    \Omega(t) =& \sum_{j=L, R} \sum_{m,n=1}^{2} \frac{\mathcal{E}_m(t)\mathcal{E}_{n}(t)}{2\Delta} e^{\iu \qty[\qty(\vb{k}_m - \vb{k}_n)\cdot\vb{r} - \qty(\phi_m-\phi_n)]} \qty(\vb{d}_{1j} \cdot \bm{\epsilon}_m) \qty(\vb{d}_{0j} \cdot \bm{\epsilon}_n)^\ast \\
    =& -\frac{\mathcal{E}_1^2}{2\Delta} d_{1} d_{0} \sin^2{\beta_1} - \frac{\mathcal{E}_2^2}{2\Delta} d_{1} d_{0} \sin^2{\beta_2} - \frac{\mathcal{E}_1 \mathcal{E}_2}{2\Delta} e^{\iu \qty(\Delta k z - \Delta\phi + \Delta\psi)} d_{1}d_{0}  \sin{\beta_1}\sin{\beta_2} - \frac{\mathcal{E}_2 \mathcal{E}_1}{2\Delta} e^{-\iu \qty(\Delta k z - \Delta\phi + \Delta\psi)} d_{1} d_{0} \sin{\beta_2}\sin{\beta_1} \notag\\
     &+ \frac{\mathcal{E}_1^2}{2\Delta} d_{1} d_{0} \cos^2{\beta_1} + \frac{\mathcal{E}_2^2}{2\Delta} d_{1} d_{0} \cos^2{\beta_2} + \frac{\mathcal{E}_1 \mathcal{E}_2}{2\Delta} e^{\iu \qty(\Delta k z - \Delta\phi)} d_{1}d_{0}  \cos{\beta_1}\cos{\beta_2} + \frac{\mathcal{E}_1 \mathcal{E}_2}{2\Delta} e^{-\iu \qty(\Delta k z - \Delta\phi)} d_{1} d_{0} \cos{\beta_2}\cos{\beta_1} \notag\\
    =& \frac{\mathcal{E}_1^2}{2\Delta} d_{1} d_{0} \cos{2\beta_1} + \frac{\mathcal{E}_2^2}{2\Delta} d_{1} d_{0} \cos{2\beta_2} + 2 \frac{\mathcal{E}_1 \mathcal{E}_2}{2\Delta} d_{1}d_{0}  \qty[\cos\beta_1 \cos\beta_2 \cos\qty(\Delta k z - \Delta\phi) -  \sin\beta_1 \sin\beta_2 \cos\qty(\Delta k z - \Delta\phi + \Delta\psi)]\notag\\
\label{eq: Rabi_collimated_with_psi}
     =& \Omega_1(t) \cos{2\beta_1} + \Omega_2(t) \cos{2\beta_2} + 2\sqrt{\Omega_1(t)\Omega_2(t)} A \cos\qty[\Delta k z - \Delta\phi(t) - \gamma],
\end{align}
where
\begin{align}
    \Omega_1(t) &= \frac{\mathcal{E}_1^2}{2\Delta} d_{1} d_{0},\\
    \Omega_2(t) &= \frac{\mathcal{E}_2^2}{2\Delta} d_{1} d_{0},\\
    \Delta k &= \qty(k_1)_z - \qty(k_2)_z,\\
    \Delta \phi(t) &= \phi_1(t) - \phi_2(t),\\
    \Delta \psi &= \psi_1 - \psi_2.
\end{align}
For convenience we have also introduced
\begin{align}
    A &= \sqrt{a^2 + b^2},\\
    \gamma &= \atantwo \qty(a, b),\\
    a &= \sin\beta_1 \sin\beta_2 \sin\Delta\psi,\\
    b &= \cos\beta_1 \cos\beta_2 - \sin\beta_1\sin\beta_2\cos\Delta\psi.
\end{align}
In Eq.~\eqref{eq: Rabi_collimated_with_psi} we have assumed $\Omega_1$ and $\Omega_2$ to be nonnegative without loss of generality. By definition~$\Omega_1$ and~$\Omega_2$ are the Rabi frequencies of each of the beams if their polarizations match the corresponding levels perfectly. Notice that the phase difference $\Delta\phi$ is time-dependent and that the difference $\Delta\psi$ in relative phases $\psi_m$'s in the polarizations show up in the phase $\gamma$ as physical observables. 
\subsubsection{\label{appx: config}Beam configuration}
Following discussions in~\cite{mizrahi_ultrafast_2013} and for simplicity, we adopt the following beam configuration.
\begin{enumerate}
    \item  
        The two beams share the same real temporal intensity envelope, so that
        \begin{align}
            \Omega_0(t) \equiv \Omega_1(t) = \Omega_2(t).
        \end{align}
    \item
        We take the beams to be counterpropagating along the quantization axis, with $\qty(k_1)_z = - \qty(k_2)_z = k$ and $\Delta k = 2k$.
    \item
        We assume a ``linear $\perp$ linear'' polarization scheme (lin $\perp$ lin), i.e.,
        \begin{align}
            \beta_1 = -\beta_2 = \pm\frac{\pi}{4},
        \end{align}
        and set the relative polarization phase to $\Delta\psi = 0$.
    \item 
        The beams share the same initial optical phase and have a Raman beat frequency~$\Delta\omega$, so that
        \begin{align}
            \Delta \phi (t) = \Delta\omega t.
        \end{align}
        By Approxs.~\ref{approx: RWA&AE_multiple_beam}, this requires
        \begin{align}
            \Delta\omega \ll \Delta,
        \end{align}        
        ensuring that the two-photon detuning is small compared to the single-photon detuning.
\end{enumerate}
Under these assumptions, the effective two-photon Rabi frequency becomes
\begin{align}
\label{eq: Rabi_final}
    \Omega(t) = 2 \Omega_0(t) \cos\qty(2 k z - \Delta\omega t).
\end{align}
\subsection{Full Hamiltonian}
\subsubsection{\label{appx: H_SP}Schr\"odinger Picture}
We now write down the full Hamiltonian of the the system. 
For a single trapped ion (\Ba, e.g.) of mass~$m$ interacting with two counterpropagating Raman laser beams configured in Sec.~\ref{appx: config}, the total Hamiltonian in the SP has the form
\begin{align}
\label{eq: H}
    H(t) = \underbrace{\frac{\vb{p}^2}{2m}}_{\mathrm{external}} + 
           \underbrace{\frac{\omegaa}{2} \sigma_z}_{\mathrm{internal}} + 
           \underbrace{\Omega_0(t) \cos\qty(2 k z - \Delta\omega t) \sigma_x}_{\mathrm{interaction}} +
           \underbrace{V(\vb{r}, t)}_{\mathrm{trap}}.
\end{align} 
Here, $\vb{r}$ and $\vb{p}$ are the position and momentum operators of the ion, and $\qty{\sigma_x, \sigma_z}$ are Pauli operators. The frequency~$\omegaa$ is the qubit transition frequency between~$\ket{0}$ and~$\ket{1}$, and~$\Omega(t)$ is the Rabi frequency given in Eq.~\eqref{eq: Rabi_final}. 

The operator $V(\vb{r}, t)$ incorporates the potential of the linear Paul trap, which has the form~\cite{leibfried_quantum_2003, nguyen_micromotion_2012}
\begin{align}
    V(\vb{r}, t) = \sum_{\mu = x, y, z}\frac{1}{8} m \omegaR^2 \mu^2 \qty[a_\mu + 2 q_\mu \cos\qty(\omegaR t + \phi_R)], 
\end{align}
where $\mu\in\qty{x, y, z}$ stands for each of the cartesian coordinates, $\omegaR$ is the RF frequency of the trap, and $\phi_R$ is the phase of the RF drive. In our case of a linear Paul trap, we have
\begin{align}
    a_z &= a_y = -\frac{1}{2} a_x,\\
    q_z &= - q_y,\quad q_x = 0.
\end{align}

We consider two counter-propagating laser beams along the~$z$ quantization axis with lin $\perp$ lin polarizations (as shown in Fig.~1(a) of the main text). The relevant trap potential is only in the $z$ direction
\begin{align}
    V(\vb{r}, t) = \frac{1}{8} m \omegaR^2 z^2 \qty[a_z + 2 q_z \cos\qty(\omegaR t + \phiR)].
\end{align}
Therefore the total Hamiltonian of interest in the SP is given by
\begin{align}
\label{eq: H_1D_SP}
    H(t) = \frac{p_z^2}{2m} + 
           \frac{1}{8} m \omegaR^2 z^2 \qty[a_z + 2 q_z \cos\qty(\omegaR t + \phiR)] +
           \frac{\omegaa}{2} \sigma_z + 
           \Omega_0(t) \cos\qty(2 k z - \Delta\omega t) \sigma_x.
\end{align}
This expression reproduces Eq.~(1) of the main text. For simplicity of notation, the main text denotes the envelope~$\Omega_0(t)$ by~$\Omega(t)$. Thus, the Rabi frequency referred to in the main text corresponds to $\Omega_0(t)$ as defined in this Supplemental Material.
\subsubsection{Interaction Picture}
Equation~\eqref{eq: H_1D_SP} can be rewritten in a form that makes the spin-motion coupling more explicit. Using the stability condition of an ion trap~\cite{nguyen_micromotion_2012}, 
\begin{align}
    a_z + \frac{q_z^2}{2} \geq 0,
\end{align}
we can rewrite the Hamiltonian by
\begin{align}
    H(t) & = \frac{p_z^2}{2m} + 
           \frac{1}{8} m \omegaR^2 z^2 \qty[a_z + \frac{q_z^2}{2} - \frac{q_z^2}{2} + 2 q_z \cos\qty(\omegaR t + \phiR)] +
           \frac{\omegaa}{2} \sigma_z + 
           \Omega_0(t) \cos\qty(2 k z - \Delta\omega t) \sigma_x \\
         & = \frac{p_z^2}{2m} + \frac{1}{2} m \omegaS^2 z^2 + \frac{1}{8} m \omegaR^2 z^2 \qty[2 q_z \cos\qty(\omegaR t + \phiR) - \frac{q_z^2}{2}] +
             \frac{\omegaa}{2} \sigma_z + 
             \Omega_0(t) \cos\qty(2 k z - \Delta\omega t) \sigma_x \\
         & = \frac{p_z^2}{2m} + \frac{1}{2} m \omegaS^2 z^2 + \frac{1}{8} m \omegaR^2 z^2 \qty[2 q_z \cos\qty(\omegaR t + \phiR) - \frac{q_z^2}{2}] +
             \frac{\omegaa}{2} \sigma_z + 
             \Omega_0(t) \cos\qty(2 k z - \Delta\omega t) \sigma_x \\
\label{eq: H_1D_SP_SDK_manifest_fock_basis_alt}
          &= \underbrace{\omegaS a^\dagger a + \frac{\omegaa}{2} \sigma_z}_{\mathrm{free\,evolution}}  +
             \underbrace{\frac{1}{16} \frac{\omegaR^2}{\omegaS} \qty(a^\dag + a)^2\qty[2 q_z \cos\qty(\omegaR t + \phiR) - \frac{q_z^2}{2}] }_{\mathrm{micromotion}}  \notag\\
          &\quad +\frac{\Omega_0(t)}{2} \qty[ \underbrace{\qty(D_+ \sigma_+ e^{-\iu \Delta\omega t} + D_- \sigma_- e^{\iu\Delta\omega t})}_{\mathrm{forward\,kick}} + \underbrace{\qty(D_+ \sigma_- e^{ - \iu \Delta\omega t} + D_- \sigma_+ e^{ \iu\Delta\omega t})}_{\mathrm{backward\,kick}}],
\end{align}
where we have used
\begin{align}
\label{eq: omega_S_micromotion}
    \omegaS &= \frac{\sqrt{a_z + \frac{q_z^2}{2}}}{2} \omegaR, \\
    z &= \sqrt{\frac{1}{2m\omegaS}} \qty(a^\dagger + a),\\
    p_z &= \iu \sqrt{\frac{m\omegaS}{2}} \qty(a^\dagger - a),\\
    \eta &= \frac{k}{\sqrt{2m\omegaS}},\\
    D_\pm &= D(\pm 2\iu \eta) = e^{\pm 2\iu \eta \qty(a^\dagger + a)} = e^{\pm 2\iu k z}.
\end{align}

We can thus move into the IP defined by the free evolution part of the Hamiltonian, i.e.,
\begin{align}
    H_0 & = \omegaS a^\dagger a + \frac{\omegaa}{2} \sigma_z,\\
\label{eq: U0}
    U_0(t) & = e^{-\iu  \omegaS t a^\dagger a}e^{-\iu\frac{\omegaa t}{2} \sigma_z},
\end{align}
and get
\begin{align}
    \tilde{H}(t) = &\underbrace{\frac{1}{16} \frac{\omegaR^2}{\omegaS} \qty(e^{\iu \omegaS t} a^\dag + e^{-\iu \omegaS t} a)^2 \qty[2 q_z \cos\qty(\omegaR t + \phiR) - \frac{q_z^2}{2}] }_{\mathrm{micromotion}} \notag\\ 
\label{eq: H_1D_IP}
      & + \frac{\Omega_0(t)}{2} \qty[\underbrace{\qty(\tilde{D}_+ \sigma_+ e^{\iu \qty(\omegaa - \Delta\omega) t} + \tilde{D}_- \sigma_- e^{- \iu\qty(\omegaa -\Delta\omega) t})}_{\mathrm{forward\,kick}} + \underbrace{\qty(\tilde{D}_+ \sigma_- e^{-\iu \qty(\omegaa + \Delta\omega) t} + \tilde{D}_- \sigma_+ e^{\iu\qty(\omegaa + \Delta\omega) t})}_{\mathrm{backward\,kick}}],
\end{align}
where
\begin{align}
    \tilde{D}_\pm(t) &= D(\pm 2\iu \eta e^{\iu \omegaS t}) = e^{\pm 2\iu \eta \qty(e^{\iu \omegaS t}a^\dagger + e^{- \iu \omegaS t} a)}.
\end{align}
This gives rise to Eq.~(2) in the main text.
\section{\label{appx: error bound}Error bound}
In this section we calculate the error bound plotted as the gray dashed line in Fig.~2(a)--(b) of the main text. We ignore the micromotion term in Eq.~\eqref{eq: H_1D_IP} for this derivation.

For a constant-amplitude Rabi frequency, we let
\begin{align}
    \Omega_0(t) = \frac{\theta}{\tau},
\end{align}
where $\theta$ is the pulse area, and $\tau$ is the SDK duration. Then the Hamiltonian becomes
\begin{align}
    \tilde{H}(t) &= \frac{\theta}{2\tau} \qty[\underbrace{\qty(D_+    \sigma_+ e^{\iu \qty(\omegaa - \Delta\omega) t} + D_- \sigma_- e^{- \iu\qty(\omegaa - \Delta\omega) t})}_{\mathrm{forward\,kick}} + \underbrace{\qty(D_+ \sigma_- e^{-\iu \qty(\omegaa + \Delta\omega) t} + D_- \sigma_+ e^{\iu\qty(\omegaa + \Delta\omega) t})}_{\mathrm{backward\,kick}}] \\
\label{eq: H_CW}
    &= g_0 \qty[\underbrace{\qty(D_+ \sigma_+ e^{\iu \omegaminus t} + D_- \sigma_- e^{- \iu\omegaminus t})}_{\mathrm{forward\,kick}} + \underbrace{\qty(D_+ \sigma_- e^{-\iu \omegaplus t} + D_- \sigma_+ e^{\iu \omegaplus t})}_{\mathrm{backward\,kick}}],
\end{align}
where we have defined
\begin{align}
    \omega_\pm &= \omega_a \pm \Delta\omega,\\
    g_0 &= \frac{\theta}{2\tau}.
\end{align}
The error bound we want to calculate is the contribution from the backward kick. Suppose the initial state is $\ket{0, 0}$. Then up to the first order in perturbation theory, the amplitude due to the backward kick is given by
\begin{align}
    c_\mathrm{back} &\approx -\iu \int_0^\tau \dt g_0 e^{\iu \omegaplus t} \notag\\
    &= -\iu g_0 \frac{e^{\iu \omegaplus \tau} - 1}{\omegaplus} \notag\\
    &= -\iu g_0 e^{\iu\omegaplus\tau/2} \frac{2\sin\frac{\omegaplus\tau}{2}}{\omegaplus}.
\end{align}
Then the upper error bound is given by
\begin{align}
    \epsilon &= |c_\mathrm{back}|^2 \leq \left|\frac{2g_0}{\omegaplus} \right|^2.
\end{align}
Under the resonant condition, we have 
\begin{align}
    \omegaa = \Delta\omega,
\end{align}
and thus
\begin{align}
    \omegaplus &= 2\omegaa, \\
    \omegaminus &= 0.
\end{align}
The error bound is therefore
\begin{align}
    \epsilon \leq \left|\frac{g_0}{\omegaa} \right|^2 = \left|\frac{\theta}{2\omegaa\tau} \right|^2.
\end{align}

\section{\label{appx: square CW} Constant-amplitude CW SDK}
In this section we derive an analytical expression for the time evolution of the system under a constant-amplitude Rabi frequency. This solution is used to generate the solid curves in Fig.~2(a) of the main text. 

We use the same Hamiltonian~\eqref{eq: H_CW} in Sec.~\ref{appx: error bound}. This Hamiltonian does not commute with itself at different times. Multiple approaches can be used to solve for an analytical solution exactly, including a Floquet approach or a Wei-Norman approach. Here we solve the Schrödinger equation by truncating the Hilbert space into subspaces of interest. We use the initial condition~$\ket{0,0}$.

By construction, a complete basis of the Hilbert space is
\begin{equation}
    \qty{\ket{0, 0}, \ket{1, \pm 2\iu \eta}, \ket{0, \pm 4\iu \eta}, \cdots},
\end{equation}
where we have assumed the parity selection rules inherent in the Hamiltonian. Taking the initial state as~$\ket{0,0}$ and the forward-kicked (target) state as~$\ket{1, 2\iu \eta}$, we make the assumption that the major source of error of our system comes from leakage into the initial state and the backward-kick state $\ket{1, 2\iu \eta}$. This allows us to solve for the forward fidelity in the manifold
\begin{align}
    \qty{\ket{0, 0}, \ket{1, \pm 2\iu \eta}}.
\end{align}
We can extend the dimension of the solution manifold to higher numbers (including the $\ket{0, \pm 4\iu \eta}$ states and further) later. 

\begin{enumerate}
    \item 
        In the three-dimensional manifold, the state vector is expressed as
        \begin{align}
            \ket{\psi(t)} = c_0(t) \ket{0, 0} + c_1^+(t) \ket{1, 2\iu \eta} + c_1^-(t) \ket{1, - 2\iu \eta}.
        \end{align}
        and the Hamiltonian is given by
        \begin{align}
            \tilde{H}(t) = g_0 \mqty*(0 & e^{-\iu \omegaminus t} & e^{-\iu \omegaplus t} \\ 
                                                    e^{\iu \omegaminus t} & 0 & 0 \\
                                                    e^{\iu \omegaplus t} & 0 & 0).
        \end{align}
        The Schrödinger equation is thus
        \begin{align}
            \iu \difft{\ket{\psi(t)}} = \tilde{H}(t) \ket{\psi(t)},
        \end{align}
        or
        \begin{align}
            \iu \difft{c_0} &= g_0 \qty(e^{-\iu \omegaminus t} c_1^+ + e^{-\iu \omegaplus t} c_1^-),\\
            \iu \difft{c_1^+} &= g_0 e^{\iu \omegaminus t} c_0,\\
            \iu \difft{c_1^-} &= g_0 e^{\iu \omegaplus t} c_0,
        \end{align}
        with the initial condition $\ket{i} = \mqty*(1\\0\\0)$.
        
        Multiple approaches can be used to solve for the three-level time-dependent Hamiltonian, including a standard Laplace transform approach or an $\SU{3}$ approach. Here we use the trick of gauge transformation to transform the Hamiltonian into a time-independent one.
        
        Consider the following gauge transformation
        \begin{align}
            b_0 &= c_0,\\
            b_1^+ &= e^{-\iu \omegaminus t} c_1^+,\\
            b_1^- &= e^{-\iu \omegaplus t} c_1^-,
        \end{align}
        or
        \begin{align}
            \ket{b} = \mqty*(b_0\\b_1^+\\b_1^-) = \mqty*(1&&\\&e^{-\iu \omegaminus t}&\\&&e^{-\iu \omegaplus t})\mqty*(c_0\\c_1^+\\c_1^-) = V(t) \mqty*(c_0\\c_1^+\\c_1^-) = V\ket{\psi}.
        \end{align}
        which leads to Schrödinger equation in the gauge transformation
        \begin{align}
            \iu \difft{\ket{b}} = M \ket{b},
        \end{align}
        with the initial condition $\ket{i}$, where $M$ is time-independent and is given by
        \begin{align}
        \label{eq: M_gauge}
            M = \iu \difft{V} V^{-1} + V H V^{-1} = \mqty*(0 & g_0 & g_0 \\ 
                                                    g_0 & \omegaminus & 0 \\
                                                    g_0 & 0 & \omegaplus).
        \end{align}
        Hence we have
        \begin{align}
            \ket{b(t)} = e^{-\iu M t} \ket{i},
        \end{align}
        and thus
        \begin{align}
            \ket{\psi(t)} = V(t)^{-1} e^{- \iu M t} \ket{i}.
        \end{align}  
        
    \item
        In the five-dimensional manifold, the state vector is expressed as
        \begin{align}
            \ket{\psi(t)} &= c_0(t) \ket{0, 0} + c_1^+(t) \ket{1, 2\iu \eta} + c_1^-(t) \ket{1, - 2\iu \eta} + c_2^+(t) \ket{0, 4\iu \eta} + c_2^-(t) \ket{0, - 4\iu \eta},
        \end{align}
        and the Hamiltonian is given by
        \begin{align}
            \tilde{H}(t) = g_0 \mqty*(0 & e^{-\iu \omegaminus t} & e^{-\iu \omegaplus t} & 0 & 0\\ 
                                                    e^{\iu \omegaminus t} & 0 & 0 & e^{\iu \omegaplus t} & 0 \\
                                                    e^{\iu \omegaplus t} & 0 & 0 & 0 & e^{\iu \omegaminus t} \\
                                                    0 & e^{-\iu \omegaplus t} & 0 & 0 & 0\\
                                                    0 & 0 & e^{-\iu \omegaminus t} & 0 & 0).
        \end{align}
        We still want to find a gauge transformation to make $\tilde{H}(t)$ time independent. This can indeed be achieved. Define the matrix elements of $V$ as
        \begin{align}
            V_{jk} = \delta_{jk} e^{\iu \chi_j t}.
        \end{align}
        Then using Eq.~\eqref{eq: M_gauge}, we have
        \begin{align}
            M_{jk} = - \delta_{jk} \chi_j + H_{jk} e^{\iu \qty(\chi_j - \chi_k)  t}.
        \end{align}
        For an $n$-dimensional $\tilde{H}$, there are $(n-1)$ phases to match and $n$ variables to define for the gauge transformation. Thus the solution is overdetermined. For convenience we choose
        \begin{align}
            \chi_{\ket{0, 2\iu m \eta}} &= m \Delta\omega,\\
            \chi_{\ket{1, 2\iu m \eta}} &= - \omegaa + m \Delta\omega.
        \end{align}
        Setting $n = 5$, we have
        \begin{align}
            V(t) = \mqty*(1&&&&\\&e^{-\iu \omegaminus t}&&&\\&&e^{-\iu \omegaplus t}&&\\&&&e^{ \iu 2\Delta\omega t}&\\&&&&e^{- \iu 2\Delta\omega t}),
        \end{align}
        and
        \begin{align}
            M = \mqty*(0 & g_0 & g_0 & 0 & 0\\ 
                g_0 & \omegaminus & 0 & g_0 & 0 \\
                g_0 & 0 & \omegaplus & 0 & g_0 \\
                0 & g_0 & 0 & -2\Delta\omega & 0\\
                0 & 0 & g_0 & 0 & 2\Delta\omega).
        \end{align}
        Similarly we have
        \begin{align}
            \ket{\psi(t)} = V(t)^{-1} e^{- \iu M t} \ket{i}.
        \end{align}  
\end{enumerate}
\section{\label{appx: micromotion}Condition for vanishing micromotion}
In this section we derive the condition for vanishing micromotion effects given in Eq.~(4) of the main text.

We begin by analyzing the system’s evolution under pure micromotion.  Because the micromotion term in Eq.~\eqref{eq: H_1D_IP} acts only on the motional degree of freedom and leaves the internal state unchanged, we may take the initial state to be purely motional, i.e.,
\begin{align}
    \ket{\psi_0} = \ket{\alpha}.
\end{align}
Since the micromotion term always commutes with other terms in Eq.~\eqref{eq: H_1D_IP}, we can treat the micromotion dynamics independently and determine the conditions under which its overall effect reduces to the identity. The Hamiltonian we study is thus
\begin{align}
\label{eq: H_mictomotion}
    \tilde{H}(t) = \frac{1}{16} \frac{\omegaR^2}{\omegaS} \qty(e^{\iu \omegaS t} a^\dag + e^{-\iu \omegaS t} a)^2 \qty[2 q_z \cos\qty(\omegaR t + \phiR) - \frac{q_z^2}{2}].
\end{align}

We now solve for the evolution of the motional state over a typical SDK time duration $\tau$. It is convenient to work in the regime
\begin{align}
    \omegaS \tau \ll 1.
\end{align}
so that the secular motion can be regarded as frozen over the pulse interval. In this limit, the micromotion term effectively commute with itself at different times and allows for a simple analytical solution.

In reality, we use
\begin{align}
    \omegaR &= 2\pi k \times 40 \mathrm{MHz}, \quad k\in\qty[0.1, 1], \\
    a_z &= 0,\\
    q_z &= 0.15,\\
    \tau &\approx 5 \mathrm{ns}.
\end{align}
This results in
\begin{align}
    \omegaS \tau = \frac{\sqrt{a_z + \frac{q_z^2}{2}}}{2} \omegaR \tau = \frac{q_z}{2\sqrt{2}} \omegaR \tau \approx \frac{k}{15} \ll 1.
\end{align}
Therefore we can ignore the secular coefficients in the Hamiltonian~\eqref{eq: H_mictomotion} and get
\begin{align}
\label{eq: H_mictomotion_approx}
    \tilde{H}(t) \approx \frac{1}{16} \frac{\omegaR^2}{\omegaS} \qty(a^\dag + a)^2 \qty[2 q_z \cos\qty(\omegaR t + \phiR) - \frac{q_z^2}{2}].
\end{align}
As argued above, the time-dependent Hamiltonian given in Eq.~\eqref{eq: H_mictomotion_approx} commutes with itself at different times and can be solved exactly, yielding
\begin{align}
    U(t_0 + \tau, t_0) &= \mathcal{T} \exp\qty[-\iu \int_{t_0}^{t_0 + \tau} \dd{t} \tilde{H}(t)] \notag\\
              &= \exp\qty[-\frac{\iu}{16} \frac{\omegaR^2}{\omegaS} \qty(a^\dag + a)^2 \int_{t_0}^{t_0 + \tau} \dd{t} \qty[2 q_z \cos\qty(\omegaR t + \phiR) - \frac{q_z^2}{2}]] \notag\\
              &= \exp\qty[-\frac{\iu}{16} \frac{\omegaR^2}{\omegaS} \qty(a^\dag + a)^2 \frac{2q_z}{\omegaR} \theta],
\end{align}
where the phase $\theta$ is given by
\begin{align}
    \theta &= \sin\qty[\omegaR\qty(t_0 + \tau) + \phiR] - \sin\qty(\omegaR t_0 + \phiR) - \frac{\omegaR \tau q_z }{4} \notag\\
      &\approx \sin\qty[\omegaR\qty(t_0 + \tau) + \phiR] - \sin\qty(\omegaR t_0 + \phiR) \notag\\
      &= 2 \cos\qty(\omegaR t_0 + \phiR + \frac{\omegaR \tau}{2}) \sin\frac{\omegaR \tau}{2},
\end{align}
where we have used for our choice of parameters,
\begin{align}
    \frac{\omegaR \tau q_z }{4} \approx \frac{k}{20} \ll 1,
\end{align}
and the relation
\begin{align}
    \sin{x} - \sin{y} = 2 \cos\frac{x + y}{2} \sin\frac{x - y}{2}.
\end{align}
Clearly, the time evolution of micromotion is trivial (identity) if we set
\begin{align}
    \theta = 0,
\end{align}
which yields
\begin{align}
\label{eq: condition_micromotion_null}
    \omegaR \qty(t_0 + \frac{\tau}{2}) + \phiR = \qty(2n + 1)\frac{\pi}{2}, \quad n \in \integers.
\end{align}
If we choose symmetric time labels such that
\begin{align}
    t_ 0 = - \frac{\tau}{2},
\end{align}
we have the condition for trivial micromotion
\begin{align}
\label{eq: condition_micromotion_null_symmetric}
    \phiR = \qty(2n + 1)\frac{\pi}{2}, \quad n \in \integers.
\end{align}

\bibliographystyle{apsrev4-2}
\bibliography{main2}